\documentclass[aps,showpacs,onecolumn,prl,11pt]{revtex4-1}
\usepackage{graphicx}
\usepackage{amsmath, amsthm, amssymb}
\usepackage[caption=false]{subfig} % Finally figures out how to make captions look right!
\newcommand*{\HOLU}{{\sc HOMO-LUMO} }
\newcommand*{\LU}{{\sc LUMO} }
\newcommand*{\HO}{{\sc HOMO} }

\newcommand*{\Tr}[1]{\operatorname{Tr}\left\{ #1 \right\}}

\newcommand*{\ve}{\varepsilon}

\newcommand*{\myT}{{\bf T}}

\graphicspath{{./Figures/}}

\begin{document}

%% The title of the article is given with the usual \title command.
%%
%% Each author should be given as a separate \author command.
%%
%% For corresponding authors please use \author* and give the email
%% address as a second mandatory argument.
%%
%% The affiliation of authors is given after the authors; the
%% affiliations are numbered consecutively.
%%
%% If some authors have the same affiliation you can use the optional
%% argument of \author and \author* to give the number of that
%% The whole block is printed with the \maketitle command at the very
%% end.
%%%%%%%%%%%%%%%%%%%%%%%%%%%%%%%%%%%%%%%%%%%%%%%%%%%%%%%%%%%%%%%%%%%%%
%\title{Many-body Treatment of Quantum Transport in Multi-channel Molecular Junction Ensembles}
\title{Transmission eigenvalue distributions in highly-conductive molecular junctions}

\author{Justin P. Bergfield}
%\author{J. P. Bergfield}
\affiliation{Departments of Chemistry and Physics, University of California, Irvine, California 92697, USA}
\email{jbergfie@uci.edu}

\author{Joshua D. Barr}
\affiliation{Department of Physics, University of Arizona, 1118 East Fourth Street, Tucson, AZ 85721}

\author{Charles A. Stafford}
\affiliation{Department of Physics, University of Arizona, 1118 East Fourth Street, Tucson, AZ 85721}
%\author[2]{Charles A. Stafford}

\begin{abstract}
{\bf Background:}
The transport through a quantum-scale device may be characterized by the transmission eigenvalues.  These values constitute a junction PIN code where, for example, in single-atom metallic contacts the number of transmission channels is also the chemical valence of the
atom.  Recently, highly conductive single-molecule
junctions (SMJ) with multiple transport
channels have been formed from benzene
molecules between Pt electrodes.  Transport through these multi-channel SMJs is a probe of both the bonding properties at the lead-molecule interface and of the molecular symmetry.
{\bf \\ Results:}
Here we utilize a many-body theory that properly describes the complementary nature of the charge carrier to calculate transport distributions through Pt-benzene-Pt junctions. We develop an effective field theory of interacting pi-electrons to accurately model the electrostatic influence of the leads and an ab initio tunneling model to describe the lead-molecule bonding. With this state-of-the-art many-body technique we calculate the transport using the full molecular spectrum and using an `isolated resonance approximation' for the molecular Green's function.
{\bf \\ Conclusions}
We confirm that the number of transmission channels in a SMJ is equal to the degeneracy of the relevant molecular orbital.  In addition, we demonstrate that the isolated resonance approximation is extremely accurate and determine that transport occurs predominantly via the \HO orbital in Pt--benzene--Pt junctions.  Finally, we show that the transport occurs in a lead-molecule coupling regime where the charge carriers are both particle-like and wave-like simultaneously, requiring a many-body description.  
\end{abstract}

%\keywords
%{\bf many-body theory; quantum transport; lead-molecule interface; effective field theory;
%multi-channel; single-molecule junction; benzene platinum junction; isolated resonance approximation; transmission eigenchannels} 
%

\maketitle

%%%%%%%%%%%%%%%%%%%%%%%%%%%%%%%%%%%%%%%%%%%%%%%%%%%%%%%%%%%%%%%%%%%%%
%% Keywords can be given with the \keywords command which takes five
%% arguments. The arguments have to be sorted.
%%%%%%%%%%%%%%%%%%%%%%%%%%%%%%%%%%%%%%%%%%%%%%%%%%%%%%%%%%%%%%%%%%%%%
%\keywords{many-body;transport;effective field theory;multi-channel;benzene platinum junction; single molecule junction}

%%%%%%%%%%%%%%%%%%%%%%%%%%%%%%%%%%%%%%%%%%%%%%%%%%%%%%%%%%%%%%%%%%%%%
%% The main text starts right here. For each required and optional
%% section of the chosen document type a special command is defined.
%%
%% It is strongly recommended to use BibTeX for managing references.
%% Citations and citation lists can be given with the \cite command.
%% Please note, that not all references have been added to the
%% example document.
%%
%% For references in floats \cite is locally redefined to
%% adds the reference to the end of the list of references.
%%%%%%%%%%%%%%%%%%%%%%%%%%%%%%%%%%%%%%%%%%%%%%%%%%%%%%%%%%%%%%%%%%%%%

% For nano article we must have Introductiona and Results and Discussion
\section{Introduction}

%Recently, highly-conductive single-molecule junctions (SMJ) with multiple
%transport channels have been formed from benzene molecules between Pt electrodes \cite{Kiguchi08}.  
%
%
%  
%The number of transmission channels for a single-atom contact between two metallic electrodes is simply given by the chemical valence of the atom 
%
%\cite{Scheer98}.
%We find that the number of dominant transmission
%channels in a single-molecule junction (SMJ) is equal to the degeneracy of the
%molecular orbital closest to the metal Fermi level \cite{Bergfield11a}.

%%%%%%

% Previous work's results
%(TODO: somehow mix with next paragraph)
%The number of transmission channels for a single-atom contact between two metallic electrodes is simply given by the chemical valence of the atom \cite{Scheer98}.  
%Recently \cite{Bergfield11a} it has been determined that the number of 
%dominant transmission channels in a single-molecule junction (SMJ) is equal 
%to the degeneracy of the molecular orbital closest to the metal Fermi level \cite{Bergfield11a}.  In this article, we focus on ensembles highly conductive Pt--benzene--Pt junctions \cite{Kiguchi08} in which the lead and molecule are in direct contact.  
%% with multiple covalent bonds.  %overlap between benzene's $\pi$-electron system and the atomic-like wavefunctions of the atomically-sharp electrode are relevant.
% 

%(TODO: I don't like this here, ideas?)
The number of transmission channels for a single-atom contact between two metallic electrodes is simply given by the chemical valence of the atom \cite{Scheer98}.  
Recently \cite{Bergfield11a} it has been determined that the number of 
dominant transmission channels in a single-molecule junction (SMJ) is equal 
to the degeneracy of the molecular orbital \cite{footnote_HOMOLUMO} closest to the metal Fermi level \cite{Bergfield11a}.  In this article, we focus on ensembles highly conductive Pt--benzene--Pt junctions \cite{Kiguchi08} in which the lead and molecule are in direct contact.
  
For a two-terminal single-molecule junction (SMJ),
the transmission eigenvalues
$\tau_n$ are eigenvalues of the elastic transmission matrix \cite{Datta95}
\begin{equation}
\myT(E)=\Gamma^{\rm L}(E) G(E) \Gamma^{\rm R}(E) G^\dagger(E),
\label{eq:trans_matrix}
\end{equation}
where $G$ is the retarded Green's function \cite{Bergfield09} of the SMJ, $\Gamma^\alpha$ is the tunneling-width matrix describing
the bonding of the molecule to lead $\alpha$ and the total transmission function $T(E) = \Tr{\myT(E)}$.
The number of transmission
channels is equal to the rank of the matrix (1), which is in turn limited by the ranks of the matrices $G$ and $\Gamma^\alpha$ \cite{Bergfield11a}.  
%Consequently, any accurate description of transport requires an accurate description of these quantities.
%that these quantities  proper description of 
The additional two-fold spin degeneracy of each resonance is considered implicit throughout this work.

%In order to properly describe the transport in a SMJ both the $G$ and $\Gamma^\alpha$ must

%must be approximated in general and can be calculated using a number of methods

% Our great theory
As indicated by Eq.~\ref{eq:trans_matrix}, any accurate description of transport requires an accurate description of $G$, which can be calculated using either single-particle or many-body methods.  %. % and $\Gamma^\alpha$.  
%The Green's function $G$ of a SMJ can be found using either (effective) single-particle or many-body theories.
In effective single-particle theories, including current implementations of density functional theory (DFT), it is often necessary \cite{Toher05,Burke06,Datta06,Geskin09} to describe the transport problem by considering an ``extended molecule,'' composed of the molecule and several 
electrode atoms.  Although this %artificial partitioning 
procedure is required in order to describe transport at all, it make it difficult, if not impossible, to assign transmission eigenchannels to individual 
molecular resonances since the extended molecule's Green's function bears little resemblance to the molecular Green's function. 
 
We utilize a nonequilibrium many-body theory based on the molecular Dyson equation (MDE) \cite{Bergfield09} to investigate transport distributions 
of SMJ ensembles.  Our MDE theory correctly accounts for wave-particle duality of the charge carriers, simultaneously 
reproducing the key features of both the Coulomb blockade and coherent transport regimes, 
alleviating the necessity of constructing an ``extended molecule.'' Consequently, we can unambiguously assign transmission eigenchannels to molecular 
resonances.  Conversely, we can also construct a junction's Green's function with only a single molecular resonance.
The theory and efficacy of this `isolated resonance approximation' are %predicting transport is 
investigated in detail. % in this article.

% in the `isolated resonance approximation,' where we only include a single %molecular resonance.  
%  we consider the efficacy of such an approach for calculating transport.

% Josh's \pi-EFT
Previous applications of our MDE theory \cite{Bergfield09,Bergfield10b,Bergfield11b} %of our many-body theory 
to transport through SMJs utilized a semi-empirical Hamiltonian \cite{Castleton02}
for the $\pi$-electrons, which accurately describes the gas-phase spectra of conjugated organic molecules.  Although this approach 
should be adequate to describe molecules weakly coupled to metal electrodes, e.g.\ via thiol linkages, in junctions
where the $\pi$-electrons bind directly to the metal electrodes \cite{Kiguchi08}, 
the lead-molecule coupling may be so strong that the molecule itself is significantly
altered, necessitating a more fundamental molecular model.

To address this issue, we have developed an {\em effective field theory of interacting $\pi$-electrons} ($\pi$-EFT),  %\cite{JoshUnpublished}
in which the form of the molecular Hamiltonian is derived from symmetry principles and electromagnetic theory (multipole expansion).
The resulting formalism constitutes a state-of-the-art many-body theory that provides a realistic description of lead-molecule
hybridization and van der Waals coupling, as well as the screening of intramolecular interactions by the metal electrodes, all of
which are essential for a quantitative description of strongly-coupled SMJs \cite{Kiguchi08}.

% Lead bonding

The bonding between the tip of electrode $\alpha$ with the molecule is characterized by the tunneling-width matrix $\Gamma^\alpha$, where the rank of 
$\Gamma^\alpha$ is equal to the number of covalent bonds formed between the two.  For example, in a SMJ where a Au electrode bonds to an organic molecule 
via a thiol group, only a single bond is formed, and there
is only one non-negligible transmission channel \cite{Djukic06,Solomon06b}.  
In Pt--benzene--Pt junctions, however, each Pt electrode forms multiple bonds
to the benzene molecule  and multiple transmission channels are observed \cite{Kiguchi08}.  
In such highly-conductive SMJs the lead and molecule are in 
direct contact and the overlap between the $\pi$-electron system of the molecule and {\em all} of the atomic-like wavefunctions of the atomically-sharp 
electrode are relevant. For each Pt tip, we consider one $s$, three $p$ and five $d$ orbitals in our calculations, which represent the evanescent tunneling 
modes in free space outside the apex atom of the tip.  
%This `atomistic' model of the lead naturally leads to multichannel transport  
This physical model for the leads accurately describes the bonding over a wide range of junction configurations. %(TODO: and is therefore great)
%Transport distribution caclculations based on this `atomistic' lead model 

In the next section, we outline the relevant aspects of our MDE theory and derive transport equations in the isolated resonance approximation.  %, which are used in the calculations presented in the results section.  
We then outline the details of our atomistic lead-molecule coupling approach, in which the electrostatic influence of the leads is treated using $\pi$-EFT and the multi-orbital lead-molecule bonding is described using an atomistic model of the electrode tip.  
%In the first section we investigate the screening and van der Waals effects when a benzene moleucle is contacted by Pt electrodes and in the second we investigate the $\Gamma$ matrix derived from an atomistic physical theory.  
Finally, the transport distributions for an ensemble of Pt--benzene--Pt junctions are shown using both the full molecular Green's function and using the isolated resonance approximation.

%,
%rat

\section{Many-body theory of transport}
%\label{sec:many_body}

%When macroscopic leads are bonded to a single molecule, a
%SMJ is formed, transforming the few-body molecular problem into a true 
%many-body problem. 

%When macroscopic leads are attached to a single molecule, a
%SMJ is formed, transforming the few-body molecular problem into a true 
%many-body problem. 
%Until recently \cite{Bergfield09}, a theory of transport in SMJs that properly accounts for both the
%{\em particle and wave character} of the electron has been lacking, so that the
%Coulomb blockade and coherent transport regimes were considered 
%`complementary' \cite{Geskin09}.  in this article, we utilize a nonequilibrium many-body theory \cite{Bergfield09} that correctly accounts for wave-particle duality, 
%reproducing the key features of both the Coulomb blockade and coherent transport regimes. 

When macroscopic leads are bonded to a single molecule, a
SMJ is formed, transforming the few-body molecular problem into a full %true 
many-body problem.  The bare molecular states are dressed by interactions with the lead electrons when the SMJ is formed, shifting and broadening them in accordance with the lead-molecule coupling.  

Until recently \cite{Bergfield09} no theory of transport in SMJs was available which properly accounted for the %, a theory of transport in SMJs that properly accounts for both the
{\em particle and wave character} of the electron, %was not available, 
so that the Coulomb blockade and coherent transport regimes were considered 
`complementary ' \cite{Geskin09}.  Here, we utilize a many-body MDE theory \cite{Bergfield09,Bergfield11b} based on %ordinary 
nonequilibrium Green's functions (NEGFs) to investigate transport in multi-channel SMJs which correctly accounts for both aspects of the charge carriers.  
%Electrostatic effects on the molecular levels due to the leads' presence will be discussed in Section \ref{sec:pi_eft}. 

In order to calculate transport quantities of interest we must determine the retarded Green's function $G(E)$ of the junction, which may be written as
\begin{equation}
	G(E) = \left[ {\bf S} E - H_{\rm mol}^{(1)} - \Sigma(E)\right]^{-1},
	\label{eq:G_full}
\end{equation}
where $H_{\rm mol}$=$H_{\rm mol}^{(1)}+H_{\rm mol}^{(2)}$ is the molecular Hamiltonian which we formally separate into one-body and two-body terms \cite{Bergfield09, Bergfield11b}.  % (see Supporting Information).  
${\bf S}$ is an overlap matrix, which in an orthonormal 
basis reduces to the identity matrix, and
\begin{equation}
	\Sigma(E) = \Sigma_{\rm T}^{\rm L}(E)+\Sigma_{\rm T}^{\rm R}(E)+\Sigma_{\rm C}(E),
\end{equation}
is the self-energy, including the effect of both a finite lead-molecule coupling via $\Sigma_{\rm T}^{L,R}$ 
and many-body interactions via the Coulomb self-energy $\Sigma_{\rm C}(E)$.  The tunneling self-energy matrices are related to the tunneling-width matrices by 
\begin{equation}
\Gamma^\alpha(E) \equiv i \left(\Sigma_{\rm T}^\alpha(E) -\left[\Sigma_{\rm T}^{\alpha}(E)\right]^\dagger\right).
%=2\pi V_{n} V^\ast_{m}\, \rho_\alpha(E),
\label{eq:Gamma_alpha}
\end{equation}
Throughout this work we shall invoke the wide-band limit in which we assume that the tunneling widths are 
energy independent $\Gamma^\alpha(E) \approx \Gamma^\alpha$.

It is useful to define a molecular Green's function $G_{\rm mol}(E) = \lim_{\Gamma^\alpha \rightarrow 0^+} G(E)$.  
In the sequential tunneling regime \cite{Bergfield09}, where lead-molecule coherences can be neglected, the molecular Green's function within MDE theory is given %exactly 
by 
\begin{equation}
	 G_{\rm mol}(E) = \left[{\bf S}E - H_{\rm mol}^{(1)} - \Sigma^{(0)}_{\rm C}(E) \right]^{-1}
	\label{eq:Gmol1} 
%	\nonumber \\
% \Rightarrow && \left[ G_{\rm{mol}}(E) \right]_{n\sigma,m\sigma } = \sum_{\nu ,\nu '} {\left[ {\cal P}(\nu ) + {\cal P}(\nu ') \right]} \frac{{\langle \nu |{d_{n\sigma} }|\nu '\rangle \langle \nu '|d_{m\sigma} ^\dag |\nu \rangle }}{{E - {E_{\nu '}} + {E_\nu } + i{0^ + }}} 
% \nonumber \\
\end{equation}
where all one-body terms are included in $H_{\rm mol}^{(1)}$ and %$G_{\rm mol} = \lim_{\Gamma^\alpha \rightarrow 0^+} G$ and 
the Coulomb self-energy $\Sigma^{(0)}$ accounts for the effect of all two-body {\em intramolecular many-body correlations exactly}.  
The full Green's function of the SMJ may then be found using the {\em molecular Dyson equation} \cite{Bergfield09} 
\begin{equation}
	G(E) = G_{\rm mol}(E) + G_{\rm mol}(E)\Delta \Sigma(E) G(E),
	\label{eq:MDE}
\end{equation}
where $\Delta \Sigma = \Sigma_{\rm T} + \Delta \Sigma_{\rm C}$ and $\Delta \Sigma_{\rm C}=\Sigma_{\rm C}-\Sigma_{\rm C}^{(0)}$.  At room temperature and for small bias voltages, $\Delta \Sigma_{\rm C}\approx 0$ in the cotunneling regime \cite{Bergfield09} 
(i.e., for nonresonant transport).  Furthermore, the inelastic transmission probability is negligible compared to the elastic transmission in that limit.    

%(TODO: Don't repeat but mention rank,etc)
The molecular Green's function $G_{\rm mol}$ is found by exactly diagonalizing the molecular Hamiltonian, including all charge states and excited states of the molecule.  Projecting onto
a basis of relevant atomic orbitals one finds
%---in the sequential-tunneling limit $\Sigma_{\rm T}\rightarrow 0$
\cite{Bergfield09,Bergfield11b}
\begin{equation}
	G_{\rm mol}(E) = \sum_{\nu, \nu'} \frac{[{\cal P}(\nu) + {\cal P}(\nu')] C(\nu,\nu')}{E-E_{\nu'}+E_{\nu}+i0^+},
	\label{eq:Gmol}
\end{equation}
where ${\cal P}(\nu)$ is the probability that the molecular state $\nu$ is occupied, $C(\nu,\nu')$ are many-body matrix elements and $H_{\rm mol} \left| \nu \right.\rangle = E_\nu \left| \nu \right.\rangle$.  In linear-response, ${\cal P}(\nu)$=$e^{-\beta \left(E_\nu - \mu N_\nu \right)}/{\cal Z}$, where ${\cal Z}$=$\sum_\nu e^{-\beta\left(E_\nu-N_\nu\mu\right)}$ is the grand canonical partition function. 

The rank-1 matrix $C(\nu,\nu')$ has elements
% in equilibrium ${\cal P}(\nu)$=$e^{-\beta(E^\nu-\mu N^\nu)}/{\cal Z}$ is the probability of the state $\nu$ being occupied with $\beta$=$1/, and 
\begin{equation}
[C(\nu,\nu')]_{n\sigma,m\sigma'} = \langle \nu | d_{n\sigma} | \nu' \rangle \langle \nu' | d^\dagger_{m\sigma'} | \nu \rangle,
	\label{eq:manybody_element}
\end{equation}
where $d_{n\sigma}$ annihilates an electron of spin $\sigma$ on the $n$th atomic orbital of the molecule and $\nu$ and $\nu'$ label molecular eigenstates with different charge.  The rank of $C(\nu,\nu')$ in conjunction with Eqs.~\ref{eq:MDE} and \ref{eq:Gmol} implies that each molecular resonance $\nu\rightarrow\nu'$ contributes at most one transmission
channel in Eq.~\ref{eq:trans_matrix}, suggesting that an $M$-fold degenerate molecular resonance could sustain a maximum of $M$ transmission channels.

\subsection{Isolated-resonance approximation}

Owing to the position of the leads' chemical potential relative to the molecular energy levels and the large charging energy of small molecules, transport in SMJs is typically dominated by individual molecular resonances.  In this subsection, we calculate the Green's function in the isolated-resonance approximation wherein only a single (non-degenerate or degenerate) molecular resonance is considered.  In addition to developing intuition and gaining insight into the  transport mechanisms in a SMJ, we also find (cf.\ Results and Discussion section) 
that the isolated-resonance approximation can be used to accurately predict the transport.

%As mentioned, the molecular Green's function may be found from the molecular Hamiltonian.
%The molecular Green's function may be expressed as \cite{Bergfield09}
%\begin{equation}
%	G_{\rm mol}(E) = \sum_{\nu,\nu'} \frac{\left[{\cal P}(\nu) + {\cal P}(\nu')\right] C(\nu,\nu')}{E-E_{\nu'} + E_\nu + i0^+},
%\end{equation}
%
%
%If the resonances of a SMJ are well separated from other resonances (relate to charging, gamma) the transport through the junction are dominated by a single resonance.  In which case complicated molecular Green's function can be accurately approximated by a single resonance.

\subsubsection{Non-degenerate molecular resonance}

If we consider a single non-degenerate molecular resonance then
\begin{equation}
	G_{\rm mol}(E) \approx \frac{\left[{\cal P}(\nu) + {\cal P}(\nu')\right] C(\nu,\nu')}{E-E_{\nu'} + E_\nu + i0^+} \equiv \frac{\tilde{\lambda} \left|\lambda \right.\rangle \langle\left. \lambda \right|}{E-\varepsilon+i0^+},
	\label{eq:Gmol_nodegen}
\end{equation}
where $\varepsilon = E_{\nu'} - E_\nu$, $C(\nu,\nu')\equiv \lambda \left|\lambda \right.\rangle \langle\left. \lambda \right|$ is the rank-1 many-body overlap
matrix and we have set $\tilde{\lambda} = [{\cal P}(\nu) + {\cal P}(\nu')]\lambda$.  
In order to solve %Dyson's equation (\label{eq:MDE}) 
for $G$ analytically, it is useful to rewrite Dyson's equation (\ref{eq:MDE}) as follows:
\begin{equation}
G(E)=\left({\bf 1} - G_{\rm mol}(E)\Delta\Sigma(E)\right)^{-1}G_{\rm mol}(E).
\end{equation}
%Applying Dyson's equation i
In the elastic cotunneling regime ($\Delta \Sigma_{\rm C}=0$) we find
\begin{eqnarray}
G(E) &=& \frac{\tilde{\lambda} \left|\lambda \right.\rangle \langle\left. \lambda \right|}{E-\varepsilon +i0^+} \left(1 + \frac{\tilde{\lambda} \langle\left. \lambda \right| \Sigma_{\rm T} \left|\lambda \right.\rangle}{E-\varepsilon +i0^+} + \right. \nonumber \\ 
&& \phantom{abdefghijklmno} \left. \left[\frac{\tilde{\lambda} \langle\left. \lambda \right| \Sigma_{\rm T} \left|\lambda \right.\rangle}{E-\varepsilon +i0^+}\right]^2 + \cdots \right) \nonumber \\ 
&=& \frac{\tilde{\lambda} \left|\lambda \right.\rangle \langle\left. \lambda \right|}{E-\varepsilon - \tilde{\lambda} \langle \lambda \left| \Sigma_{\rm T} \right| \lambda \rangle}.
\label{eq:G_isolated_dyson}
\end{eqnarray}
Equation \ref{eq:G_isolated_dyson} can equivalently be expressed as
\begin{equation}
	G(E) \approx \frac{\left[{\cal P}(\nu) + {\cal P}(\nu')\right] C(\nu,\nu')}{E-\varepsilon - \tilde{\Sigma}}, %_{\nu\nu'}},
	\label{eq:G_nodegen}
\end{equation}
where 
\begin{equation}
\tilde{\Sigma}=[{\cal P}(\nu) + {\cal P}(\nu')] \Tr{C(\nu,\nu')\Sigma_{\rm T}} 
\end{equation}
is the effective self-energy at the resonance, which includes the effect of many-body correlations via the $C(\nu,\nu')$ matrix.

Using Eq.~\ref{eq:trans_matrix}, the transmission in the isolated-resonance approximation is given by
\begin{equation}
	T(E) = \frac{\tilde{\Gamma}^{\rm L} \tilde{\Gamma}^{\rm R}}{(E-\ve)^2 + \tilde{\Gamma}^2},
	\label{eq:Isolated_trans}
\end{equation}
where 
\begin{equation}
\tilde{\Gamma}^\alpha = %\tilde{\lambda}  \langle \lambda \left| \Gamma^\alpha \right| \lambda \rangle = 
[{\cal P}(\nu) + {\cal P}(\nu')] \Tr{C(\nu,\nu')\Gamma^\alpha} 
\end{equation}
is the dressed tunneling-width matrix and $\tilde{\Gamma} = (\tilde{\Gamma}^{\rm L}+\tilde{\Gamma}^{\rm R}) /2$.

As evidenced by Eq.~\ref{eq:Isolated_trans}, the isolated-resonance approximation gives an intuitive prediction for the transport.  
Specifically, the transmission function is a single Lorentzian resonance centered about $\ve$ with a half-width at half-maximum of 
$\tilde{\Gamma}$.  The less-intuitive many-body aspect of the transport problem is encapsulated in the effective tunneling-width matrices $\tilde{\Gamma}^\alpha$, where the overlap of molecular many-body eigenstates can reduce the elements of these matrices and may strongly affect the predicted transport.

\subsubsection{Degenerate molecular resonance}

The generalization of the above results to the case of a degenerate molecular resonance is formally straightforward.  
For an $M$-fold degenerate molecular resonance
\begin{equation}
	G_{\rm mol}(E) \approx  \frac{\tilde{\lambda}}{E-\varepsilon+i0^+} \sum_{l=1}^M \left|\lambda_l \right.\rangle \langle\left. \lambda_l \right|.
	\label{eq:Gmol_degen}
\end{equation}
The $M$ degenerate eigenvectors of $G_{\rm mol}$ may be chosen to diagonalize $\Sigma_{\rm T}$ on the degenerate subspace
\begin{equation}
\tilde{\lambda}\langle \lambda_i | \Sigma_{\rm T} | \lambda_j \rangle = \delta_{ij} \tilde{\Sigma}_i
\end{equation}
and Dyson's equation may be solved as before
\begin{equation}
	G(E) \approx \sum_{l=1}^M \frac{\left[{\cal P}(\nu_l) + {\cal P}(\nu')\right] C(\nu_l,\nu')}{E-E_{\nu'}+E_{\nu_l} -\tilde{\Sigma}_{l}}.
\end{equation}
%Although the SMJ's Green's function may be found using Dyson's equation as before, the molecular eigenstates that span the degenerate sub-spaces are not orthogonal meaning that there is not generally a simple closed form expression for $G(E)$.  Moreover, 
Although $\Sigma_{\rm T}$ is diagonal in the basis of $\left| \lambda_l \right. \rangle$, $\Gamma^{\rm L}$ and $\Gamma^{\rm R}$ need not be separately diagonal.  Consequently, there is no general simple expression for $T(E)$ for the case of a degenerate resonance, but ${\bf T}$ can still be computed using Eq.~\ref{eq:trans_matrix}.

In this article we focus on transport through Pt-benzene-Pt SMJs where the relevant molecular resonances (\HO or {\sc LUMO}) are doubly degenerate.  Considering the \HO resonance of benzene
\begin{equation}
	G(E) \approx \frac{C(\nu_1,0_6)}{E-\ve_{\rm HOMO} -\tilde{\Sigma}_{\nu_1 \nu'}} +  \frac{ C(\nu_2,0_6)}{E-\ve_{\rm HOMO} -\tilde{\Sigma}_{\nu_2 \nu'}},
\label{eq:isoresapprox}
\end{equation}
where $\nu_{1,2} \in 0_5$ diagonalize $\Sigma_{\rm T}$ and $0_N$ is the $N$-particle ground state.

%(TODO: Give explanation of numerical solution to this problem?)
 
%\begin{eqnarray}
%	G(E) &\approx& \frac{\left[{\cal P}(\nu_1) + {\cal P}(\nu')\right] C(\nu_1,\nu')}{E-E_{\nu'}+E_{\nu_1} -\tilde{\Sigma}_{\nu_1 \nu'}} + \nonumber \\ 
%	&& \phantom{abcdefghijkl} \frac{\left[{\cal P}(\nu_2) + {\cal P}(\nu')\right] C(\nu_2,\nu')}{E-E_{\nu'}+E_{\nu_2} -\tilde{\Sigma}_{\nu_2 \nu'}}.
%\end{eqnarray}
%%%%%%%%%%%%%%%%%%%%%%%%%%%%%%%%%%%%%%%%%

\section{Pi-electron effective field theory}
%\label{sec:pi_eft}

In order to model the degrees of freedom most relevant
for transport, we have developed an effective field theory of
interacting $\pi$-electron systems ($\pi$-EFT) as described in detail in Ref.\citenum{Barr11}.   %TODO.
Briefly, this was done by starting with the full electronic
Hamiltonian of a conjugated organic molecule and dropping degrees
of freedom far from the $\pi$-electron energy scale.
The effective $\pi$-orbitals were then assumed to possess
azimuthal and inversion symmetry, and the effective Hamiltonian
was required to satisfy particle-hole symmetry and be explicitly local.
Such an effective field theory is preferable to semiempirical methods for applications in molecular junctions
because it is more fundamental, and hence can be readilly generalized to include screening of intramolecular Coulomb
interactions due to nearby metallic electrodes.
%(TODO: add about how much better this is than semiempircal methods)

\subsection{Effective Hamiltonian}

This allows the effective Hamiltonian for the $\pi$-electrons in gas-phase benzene to
be expressed as
\begin{eqnarray}
\label{benzeneEffectiveHamiltonian}
H_{\rm mol} & = & \mu \sum_n \rho_n - t \sum_{\langle n,m \rangle, \sigma} d^\dagger_{n\sigma} d_{m\sigma} \nonumber \\
& + &\frac{1}{2} \sum_{nm} U_{nm} (\rho_n - 1) (\rho_m - 1),
\end{eqnarray}
where $t$ is the tight-binding matrix element, $\mu$ is the
molecular chemical potential, $U_{nm}$ is the Coulomb interaction between
the electrons on the $n$th and $m$th $\pi$-orbitals, and $\rho_n \equiv \sum_\sigma d^\dagger_{n\sigma} d_{n\sigma}$.  The interaction
matrix $U_{nm}$ is calculated via a multipole expansion keeping terms up to the quadrupole-quadrupole interaction: %i.e.:
\begin{eqnarray*}
U_{nm} & = & U_{nn}\delta_{nm} \nonumber \\ 
& + & (1 - \delta_{nm})\left(U^{MM}_{nm} + U^{QM}_{nm} + U^{QM}_{mn} + U^{QQ}_{nm}\right) \nonumber \\
& + & {\cal O}(r^{-6}),
\end{eqnarray*}
where $U^{MM}$ is the monopole-monopole interaction, $U^{QM}$ is the quadrupole-monopole interaction, and $U^{QQ}$ is the quadrupole-quadrupole
interaction. For two orbitals with arbitrary quadrupole moments $Q^{ij}_n$ and $Q^{kl}_m$ separated by a displacement $\vec r$, the expressions
for these are
\begin{eqnarray*}
U^{MM}_{nm} & = & \frac{e^2}{\epsilon r}, \label{monopoleMonopole} \\
U^{QM}_{nm} & = & \frac{-e}{2 \epsilon r^3} \sum_{ij} Q_m^{ij} \hat{r}_i\hat{r}_j, \\
U^{QM}_{mn} & = & \frac{-e}{2 \epsilon r^3} \sum_{ij}  Q_n^{ij} \hat{r}_i \hat{r}_j,  \\
U^{QQ}_{nm} & = & \frac{1}{12 \epsilon r^5} \sum_{ijkl}  Q_n^{ij} Q_m^{kl} W_{ijkl}, \label{quadrupoleQuadrupole}
\end{eqnarray*}
where
\begin{eqnarray*}
W_{ijkl} & = & \delta_{li} \delta_{kj} + \delta_{ki} \delta_{lj} - 5 r^{-2}(r_k \delta_{li} r_j + r_k r_i \delta_{lj} + \delta_{ki} r_j r_l \\
& & + r_i \delta_{kj} r_l + r_k r_l\delta_{ij}) + 35 r^{-4} r_i r_j r_l r_k
\end{eqnarray*}
is a rank-4 tensor that characterizes the interaction of two quadrupoles and $\epsilon$ is a dielectric constant included to account
for the polarizability of the core and $\sigma$ electrons.  Altogether, this provides an expression for the interaction energy
that is correct up to fifth order in the interatomic distance. 

\begin{figure}
   \centering
    \includegraphics{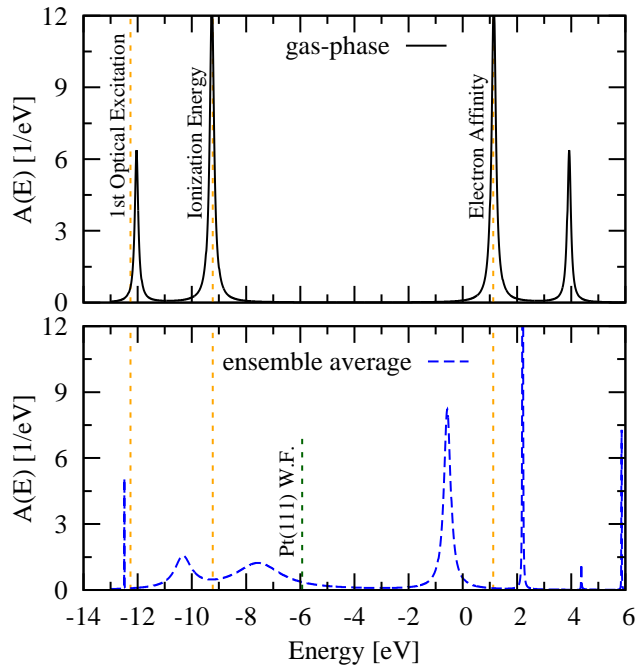}\label{fig:benzene_spectral_fcns}
\caption[]{Spectral functions $A(E)=-1/\pi \Tr{G(E)}$ at room temperature for gas-phase benzene (top panel) and 
Pt-benzene-Pt junctions (ensemble average, bottom panel).
The gas-phase resonances are broadened artificially as a guide to the eye. The dashed orange
lines are fixed by (left to right) the lowest-lying optical excitation of the molecular cation \cite{Kovac80, Sell78, Kobayoshi78, Schmidt77, Baltzer97}, 
the vertical ionization energy of the neutral molecule \cite{Howell84, Kovac80, Sell78, Kobayoshi78, Schmidt77}, and the vertical
electron affinity of the neutral molecule \cite{Burrow87}.  The asymmetry in the average spectral function arises because the 
{\sc HOMO} resonance couples more strongly on average to the Pt tip atoms than does the {\sc LUMO} resonance.  The work function of the Pt(111) surface ($-5.93$eV \cite{CRC}) is shown for reference. 
%ensemble is not completely random and consequently the particle-hole symmetry is not preserved (TODO: more?)
}
\end{figure}

\subsection{Benzene}

The adjustable parameters in our Hamiltonian for gas-phase benzene are the nearest-neighbor tight-binding
matrix element $t$, the on-site repulsion $U$, the dielectric constant $\epsilon$, and the $\pi$-orbital quadrupole moment $Q$. 
These were renormalized by fitting to experimental values that should be accurately reproduced within a $\pi$-electron only model. In particular,
we simultaneously optimized the theoretical predictions of 1)
the six lowest singlet and triplet excitations of the neutral molecule, 
2) the vertical ionization energy, 
and 3) the vertical electron affinity. The optimal parametrization for the $\pi$-EFT was found to be $t = 2.70$ eV,
$U = 9.69$ eV, $Q = -0.65$ $e${\AA}$^2$ and $\epsilon = 1.56$ with a RMS relative error of $4.2$ percent in the fit of the excitation spectrum.

The top panel of Fig.~\ref{fig:benzene_spectral_fcns} shows the spectral function for gas-phase benzene within $\pi$-EFT, along with experimental values
for the first optical excitation of the cation ($3.04$ eV), the vertical ionization energy ($9.23$ eV), and the vertical electron affinity ($-1.12$ eV).  
As a guide to the eye, the spectrum
has been broadened artificially using a tunneling-width matrix of $\Gamma_{nm} = (0.2 \; \textrm{eV}) \delta_{nm}$. The close agreement between the experimental values and the
maxima of the spectral function suggests our model is accurate at this energy scale. In particular, the accuracy of the theoretical value for the lowest optical excitation of the cation is
noteworthy, as this quantity was not fit during the renormalization procedure but rather represents a prediction of our $\pi$-EFT.

In order to incorporate screening by metallic electrodes into $\pi$-EFT, we utilized an image multipole method whereby 
the interaction between an orbital and image orbitals are included up to the quadrupole-quadrupole interaction in a screened interaction matrix $\tilde{U}_{nm}$.
In particular, we chose a symmetric $\tilde{U}_{nm}$ that ensures the Hamiltonian gives the energy required to assemble the charge distribution from
infinity with the electrodes maintained at fixed potential, namely
\[
\tilde{U}_{nm} = U_{nm} + \delta_{nm}U_{nn}^{(i)} + \frac{1}{2}(1 - \delta_{nm})(U_{nm}^{(i)} + U_{mn}^{(i)}),
\]
where $U_{nm}$ is the unscreened interaction matrix and $U_{nm}^{(i)}$ is the interaction between the $n$th orbital and the image of the $m$th orbital.
When multiple electrodes are present, the image of an orbital in one electrode produces images in the others, resulting in an
effect reminiscent of a hall of mirrors. We deal with this by including these ``higher order'' multipole moments iteratively until the
difference between successive approximations of $\tilde{U}_{nm}$ drops below a predetermined threshold.

In the particular case of the Pt--benzene--Pt junction ensemble described in the next section, %\cref{sec:lead_mol_coupling},
%(TODO: Insert section reference)
the electrodes of each junction are modeled as perfect spherical conductors.
An orbital with monopole moment $q$ and quadrupole moment $Q^{ij}$ located a distance $r$ from the center of an electrode
with radius $R$ then induces an image distribution at $\tilde{r} = \frac{R^2}{r}$ with monopole and quadrupole moments
\[
\tilde{q} =  -q \frac{R}{r} - \frac{R}{2r^3} \sum_{ij} Q^{ij} \hat{r}_i \hat{r}_j
\]
and
\[
\tilde{Q}^{ij} = -\left(\frac{R}{r}\right)^5 \sum_{kl} T_{ik} T_{jl} Q^{kl}
\]
respectively. Here $T_{ik}$ is a transformation matrix representing a reflection about the plane normal to the vector $\hat{r}$: %i.e.:
\[
T_{ik} = \delta_{ik} - 2 \hat{r}_i \hat{r}_k.
\]%(TODO: word me)
The lower panel of Fig.~\ref{fig:benzene_spectral_fcns}
 shows the Pt--benzene--Pt spectral function averaged over the ensemble of junctions described in
the next section %\cref{sec:lead_mol_coupling} 
using this method.  Comparing the spectrum with the gas-phase spectral function shown in the top panel of Fig.~\ref{fig:benzene_spectral_fcns}, we see that screening due to the nearby Pt tips reduced the
\HOLU gap by 33\% on average, from 10.39eV in gas-phase to 6.86eV over the junction ensemble.
%(TODO: Insert value) percent on average.
%% Gas gap = 10.39eV
%% Junction gap = 6.86eV

\begin{figure}
   \centering
    \includegraphics{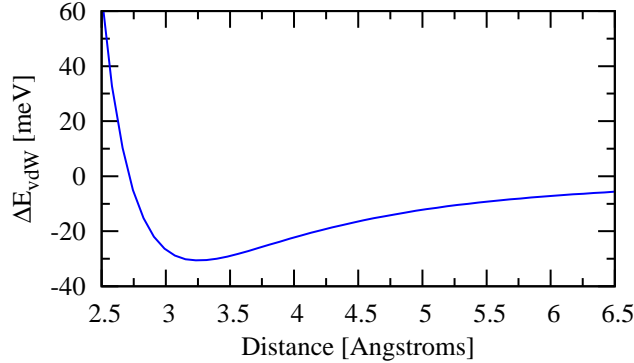}
	\caption{Calculated van der Waals contribution to the binding energy of benzene adsorbed on a Pt(111) surface as a function of distance.  Here the plane of the molecule is oriented
parallel to the Pt surface.  A phenomenological short-range repulsion $\propto r^{-12}$ has been included to model the Pauli repulsion when the $\pi$-orbitals overlap the Pt surface states.}
    \label{fig:vdW_vs_distance}
\end{figure}

The screening of intramolecular Coulomb interactions by nearby conductor(s) illustrated in Fig.~\ref{fig:benzene_spectral_fcns} leads to 
an attractive interaction between a molecule and a metal surface (van der Waals interaction).
By diagonalizing the molecular Hamiltonian with and without the effects of screening included in $U_{nm}$, 
it is possible to determine the van der Waals interaction at arbitrary temperature between a neutral molecule and a metallic electrode 
by comparing the expectation values of the Hamiltonian in these two cases:
\[
\Delta E_{vdW} = \langle H \rangle - \langle \tilde{H} \rangle
\]
This procedure was carried out at zero temperature for benzene oriented parallel to the surface of a planar Pt
electrode at a variety of distances, and the results are shown in Fig.~\ref{fig:vdW_vs_distance}.  
Note that an additional phenomenological short-range repulsion $\propto r^{-12}$ has been included in the calculation
to model the Pauli repulsion arising when the
benzene $\pi$-orbitals overlap the Pt surface states.
%Note that far from the surface of the plane $\Delta E_{vdW} \propto r^{-3}$, which is the expected result based on simple perturbation theory.  (Highlight importance of charge-charge correlations)

\section{The lead-molecule coupling}
%\label{sec:lead_mol_coupling}

When an isolated molecule is connected to electrodes and a molecular junction is formed, the energy levels of the molecule are broadened and shifted as a result of the formation of a lead-molecule bond and the electrostatic influence of the leads.  The bonding between lead $\alpha$ and the molecule is described by the tunneling width matrix $\Gamma^\alpha$ and the electrostatics, including intramolecular screening and van der Waals effects, are described by the effective molecular Hamiltonian derived using the aforementioned $\pi$-EFT.  Although we use the Pt--benzene--Pt junction as an example here, the techniques we discuss are applicable to any conjugated organic molecular junction.

%In this work, we focus on ensembles of junctions.  In a prior work \cite{Bergfield11}, the  $\Gamma^\alpha$ ensemble was generated using a random walk technique which ensured the leads had an appropriate number of channels for the valence of the atomic species.  Here we instead consider a more physical description of the leads, including all relevant orbitals of the atomic-like lead contact atom.  Focusing specifically on the Pt-benzene-Pt junctions we include the $s$, $p$ and $d$ atomic orbitals of Pt (TODO: the C value also accounts for the rest of the material..).
%
%
% was Unlike in Ref.\citenum{Bergfield11} we don't construct the $\Gamma^\alpha$ matricies randomly but instead use the following procedure.
%
% In the atop configuration the distance between the tip atom and the center of the benzene ring is 2.25\AA\ giving a tip to orbital distance of 2.65\AA\ (the C-C bonds are taken as 1.4\AA\).  The tip's position in the plane parallel to the benzene ring is a Gaussian random variable with a standard deviation of 0.5\AA\, chosen because of the preferred bonding in this region.  The height of the tip is chosen such that the closest carbon atom is 2.65\AA\.)

\begin{figure}
	\centering
	\includegraphics{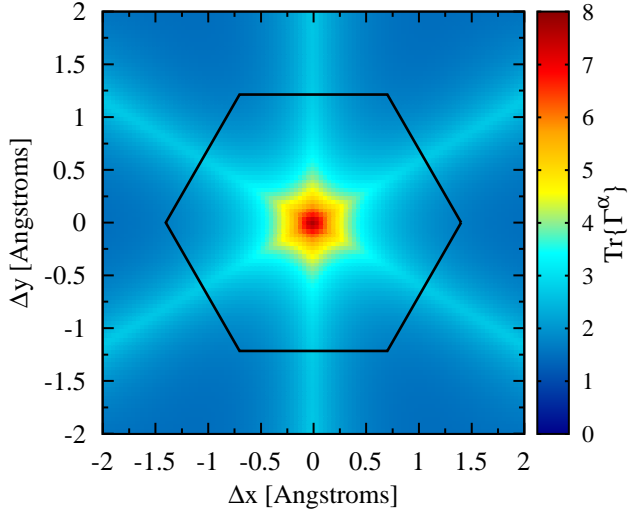}
	\caption{The trace of $\Gamma^\alpha$ for a Pt electrode in contact with a benzene molecule.  
Nine total basis states of the Pt tip are included in this calculation (one $s$, three $p$ and five $d$ states).
The tip height above the plane of the molecule is adjusted at each point such that the Pt-C distance is fixed to 2.65\AA\ (see text).  $\Tr{\Gamma^\alpha}$ retains the (six-fold) symmetry of the molecule and is sharply peaked near the center of the benzene ring indicating the strongest
bonds are formed when the lead is in the `atop' configuration.  The benzene molecule is shown schematically with the black lines; the carbons atoms are located at each vertex.}
	\label{fig:benzene_contact_image}
\end{figure}

\subsection{Bonding}

The bonding between the tip of electrode $\alpha$ with the molecule is characterized by the tunneling-width matrix $\Gamma^\alpha$ given by Eq.~\ref{eq:Gamma_alpha}.  When a highly-conductive SMJ \cite{Kiguchi08} is formed the lead and molecule are in direct contact such that the overlap between the $\pi$-electron system of the molecule and {\em all} of the atomic-like wavefunctions of the atomically-sharp electrode are relevant.  In this case we may express the elements of $\Gamma^\alpha$ as \cite{Bergfield09} %,Chen93}
%
%The tunnel coupling of the tip to atomic orbitals $n$ and $m$ of the system of interest is described by the tunneling-width matrix \cite{Bergfield09}
\begin{equation}
\Gamma^\alpha_{nm}(E) = 2\pi  \sum_{l \in \{s,p,d,\ldots \}} C_l V^{n}_l \left(V^{m}_l\right)^\ast \, \rho^\alpha_l(E),
\label{eq:Gamma_p}
\end{equation}
where the sum is over evanescent tunneling modes emanating from the metal tip, labeled by their angular momentum quantum numbers,
%all relevant atomic orbitals (and associated quantum numbers) of the tip metal, 
$\rho^\alpha_l(E)$ is the local density of states on
the apex atom of electrode $\alpha$, and $V^{n}_l$ is the tunneling matrix element of orbitals $l$ \cite{Chen93}.  
%Although we use a basis of hydrogenic orbitals, the factor $C$ is used to account for the presence of the bulk material \cite{Chen93} (TODO: more?).  
The constants $C_l$ can in principle be determined by matching the evanescent tip modes to the wavefunctions within the metal tip \cite{Chen93};
however, we set $C_l=C \, \forall \,  l$ and determine the constant $C$ by fitting to the peak of the experimental conductance histogram \cite{Kiguchi08}.
In the calculation of the matrix elements, we use the effective Bohr radius of a $\pi$-orbital
$a^* = a_0/Z$, where $a_0\approx$ 0.53{\AA} is the Bohr radius and $Z=3.22$ is the effective hydrogenic charge associated with 
the $\pi$-orbital quadrupole moment $-0.65e$\AA$^2$ determined by $\pi$-EFT. 

For each Pt tip, we include one $s$, three $p$ and five $d$ orbitals in our calculations, which represent the evanescent tunneling modes in free space
outside the apex atom of the tip.  
At room temperature, the Pt density of states (DOS) $\rho^\alpha(E) = \sum_l \rho^\alpha_l(E)$ is sharply peaked around the Fermi energy \cite{Kleber73} with $\rho^\alpha(\varepsilon_F)$=2.88/eV %, nearly 10$\times$ that of Au at the Fermi energy 
\cite{Kittel_KinderBook}.  In accordance with Ref.\ \citenum{Chen93}, we distribute the total DOS such that the $s$ orbital contributes 10\%, 
the $p$ orbitals contribute 10\%, and the $d$ orbitals contribute 80\%.

\begin{figure}
	\centering
		\includegraphics{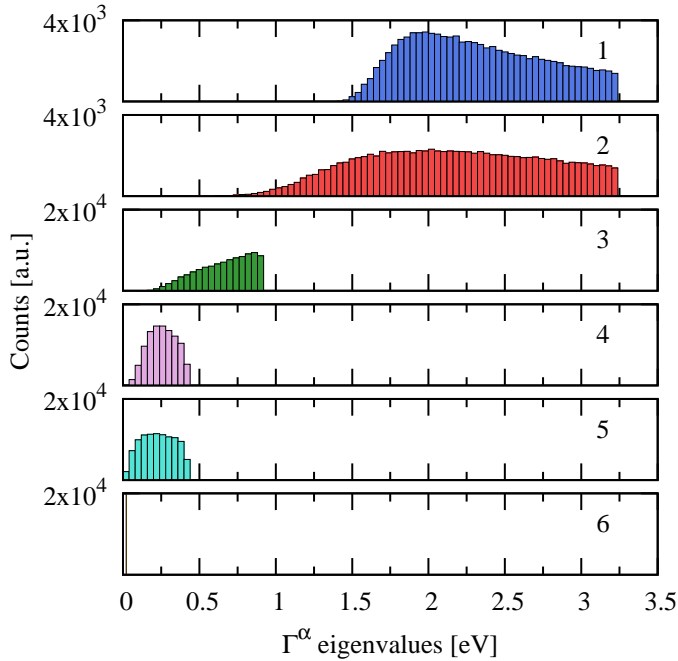}
	\caption{Eigenvalue decomposition of an ensemble of $\Gamma^\alpha$ matricies, showing that each lead-molecule contact has $\sim5$ channels. 
Note that nine orthogonal basis orbitals were included in the calculation for each lead. 
}
	\label{fig:Gamma1_eigenchannel}
\end{figure}

%(TODO: work)
We are interested in investigating transport through stable junctions where the `atop' binding configuration of benzene on Pt has the largest 
binding energy \cite{Cruz07,Morin03,Saeys02}.  In this configuration, the distance between the tip atom and the center of the benzene ring is 
$\approx$ 2.25{\AA} \cite{Kiguchi08}, giving a tip to orbital distance of $\approx$ 2.65{\AA} (the C--C bonds are taken as 1.4\AA).  
The trace of $\Gamma^\alpha(\varepsilon_F)$ is shown 
as a function of tip position in Fig.~\ref{fig:benzene_contact_image}, where for each tip position the height was adjusted such that the distance to the closest carbon atom was 
2.65\AA.  From the figure, it is evident that the lead-molecule coupling strength is peaked when the tip is in the vicinity of center of the 
benzene ring (whose outline is drawn schematically in black).  
As shown in Ref.\ \cite{Bergfield11a}, the hybridization contribution to the binding energy is %$\propto {\rm Tr}\{\Gamma\}$
\begin{eqnarray}
\Delta E_{\rm hyb} &=& \sum_{\nu\in {\cal H}_{N-1}} \int_{\mu}^\infty \frac{dE}{2\pi} \frac{\Tr{\Gamma(E)C(\nu,0_{N})}}{E-E_{0_{N}}+E_\nu} \nonumber \\
       &+&
       \sum_{\nu'\in {\cal H}_{N+1}} \int_{-\infty}^\mu \frac{dE}{2\pi} \frac{{\rm Tr} \{\Gamma(E)C(0_{N},\nu')\}}{-E-E_{0_{N}}+E_{\nu'}}, \nonumber
       \label{eq:delta_Ehybrid}
\end{eqnarray}
which is roughly $\propto {\rm Tr}\{\Gamma(\varepsilon_F)\}$. 
Here $\mu$ is the chemical potential of the lead metal, ${\cal H}_{N}$ is the $N$-particle molecular Hilbert space, 
and $0_{N}$ is the ground state of the $N$-particle manifold of the neutral molecule.
The sharply peaked nature of $\Tr{\Gamma^\alpha}$ seen in Fig.~\ref{fig:benzene_contact_image} is thus consistent with the large binding energy of the atop configuration.  

This result motivates our procedure for generating the ensemble of junctions, where we consider the tip position in the plane parallel to the benzene ring as a 2-D Gaussian random variable with a standard deviation of 0.25\AA, chosen to corresponded with the preferred bonding observed in this region.  
For each position, the height of each electrode (one placed above the plane and one below) is adjusted such that the closest carbon to the apex
atom of each electrode is 2.65\AA.  Each lead is placed independently of the other.  This procedure ensures that the full range of possible, bonded junctions are included in the ensemble.

%(TODO: refine)
The eigenvalue distributions of $\Gamma^\alpha$ over the ensemble are shown in Fig.\ref{fig:Gamma1_eigenchannel}.  
Although we include nine (orthogonal) basis orbitals for each lead, the $\Gamma$ matrix 
only exhibits five nonzero eigenvalues, presumably because only five linear combinations can be formed which are directed toward the molecule.
%does not exhibit six distinct eigenvalues because the $\pi$-orbital system of the molecule is not orthogonal to each atomic orbital.  
Although we have shown the distribution for a single lead, the number of transmission channels for two leads, 
where each $\Gamma^\alpha$ matrix has the same rank, will be the same even though the overall lead-molecule coupling strength will be larger.  
The average coupling per orbital with two electrodes is shown in the bottom panel of Fig.~\ref{fig:benzene_Unm_Trace_Gamma}.

\subsection{Screening}

%The electrostatic coupling between the molecular $\pi$-electronic system and the electrodes' metallic systems are included via the aforementioned $\pi$-EFT, where the effective molecular Hamiltonian for the $\pi$-electrons used in our many-body MDE theory is calculated using a multipole expansion for the intramolecular interactions.

The ensemble of screened interaction matrices $\tilde{U}_{nm}$
is generated using the same procedure discussed above.  Each Pt electrode is modelled as a conducting sphere with radius
equal to the Pt polarization radius (1.87\AA).  This is equivalent to the assumption that screening is due mainly to the apex atoms of each Pt tip.
The screening surface is placed such that it lies one covalent radius away from the nearest carbon atom \cite{Barr11}.
%but with the tip height in the atop configuration taken as 2.76\AA (2.25 + 1.87-1.36), slightly larger than the 2.65\AA used in the bonding ensemble.  
%The electrodes are assumed to be spheres with the polarization radius of Pt (1.87\AA).  In order for the effective screening plane of these spheres to be 
%consistent with the 2.25\AA \cite{Kiguchi08} used in the bonding ensemble it is necessary to move them by 1.87-1.36\AA, which is the difference of the 
%polarization and covalent radii of Pt (TODO: correct?).  The new tip height corresponds to a Pt-C distance of 3.095\AA, with the C-C distance taken as 
%1.4\AA.  

\begin{figure}
	\centering
		\includegraphics{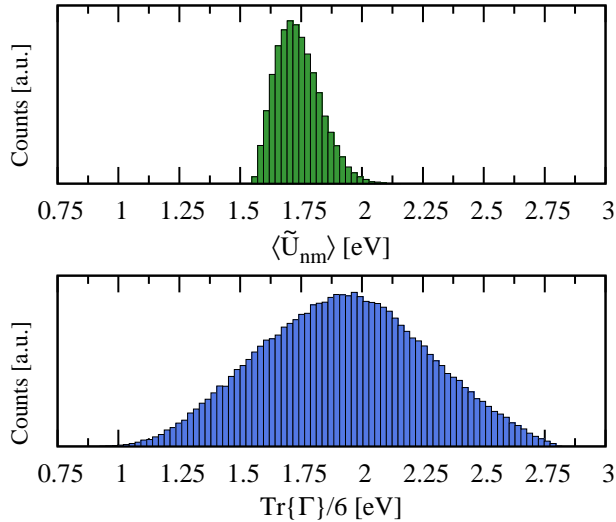}
	\caption{The distribution of charging energy $\langle \tilde{U}_{nm} \rangle$ (top panel) and $\Tr{\Gamma}$ (bottom panel) over the ensemble 
described in the text. Here $\Gamma=\Gamma^1+ \Gamma^2$ is the total tunneling-width matrix of the junction. 
The width of the $\Tr{\Gamma}$ distribution is $\sim 4\times$ that of the $\langle U_{nm} \rangle$ distribution.  
The peak of the $\langle \tilde{U}_{nm} \rangle$ and $\Tr{\Gamma}/6$ distributions are $1.58eV$ and $1.95eV$, respectively, suggesting that 
transport occurs in an intermediate regime where both the particle-like and wave-like character of the charge carriers must be considered.}
	\label{fig:benzene_Unm_Trace_Gamma}
\end{figure}

The average over the interaction matrix elements $\langle \tilde{U}_{nm} \rangle$ defines the ``charging energy'' of the molecule in the junction
\cite{Barr11}.
The charging energy $\langle \tilde{U}_{nm} \rangle$ and per orbital $\Tr{\Gamma}$ distributions are shown in the top and bottom panels of Fig.~\ref{fig:benzene_Unm_Trace_Gamma}, respectively, where two electrodes are used in all calculations.  As indicated by the figure, 
the $\Tr{\Gamma}/6$ distribution is roughly four times as broad as the charging energy distribution. % for the ensemble.  
This fact justifies using the ensemble-average $\tilde{U}_{nm}$ matrix for transport calculations \cite{Bergfield11a},
an approximation which makes the calculation of thousands of junctions computationally tractable.
The peak values of the $\langle \tilde{U}_{nm} \rangle$ and $\Tr{\Gamma}/6$ distributions are 1.58eV and 1.95eV, respectively, suggesting that
transport occurs in an intermediate regime where both the particle-like and wave-like character of the charge carriers must be considered.

In addition to sampling various
bonding configurations, we also consider the ensemble of junctions 
%produced in the experiment \cite{Kiguchi08}
to sample all possible Pt surfaces.  The work function of Pt ranges from 
5.93eV to 5.12eV for the (111) and (331) surfaces,
respectively \cite{CRC}, and we assume that $\mu_{\rm Pt}$ is 
distributed uniformly over this interval.

\begin{figure}
	\centering
		\includegraphics{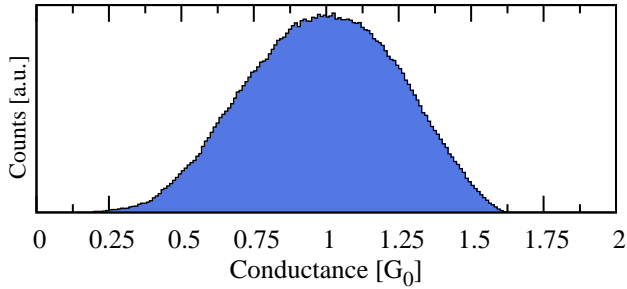}
	\caption{Calculated conductance histogram for the ensemble over bonding configurations and Pt surfaces.  The value of the conductance
peak has been fit to match the experimental data \cite{Kiguchi08}, determining the constant $C$ in Eq.~\ref{eq:Gamma_p}.  
%and width of the distribution 
There is no peak for $G \sim 0$ because we designed an ensemble of junctions where both electrodes are strongly bound to the molecule.}
	\label{fig:benzene_conductance_histogram}
\end{figure}

Using this ensemble, the conductance histogram over the ensemble of junctions can be computed, and is shown in Fig.~\ref{fig:benzene_conductance_histogram}.
%From the ensemble we can also calculate the experimentally measured conductance histrogram
The constant prefactor $C$ appearing in the tunneling matrix elements \cite{Chen93} in Eq.~\ref{eq:Gamma_p} was determined by fitting the peak
of the calculated conductance distribution to the that of the experimental conductance histogram \cite{Kiguchi08}.
Note that the width of the calculated conductance peak is also comparable to that of the experimental peak  \cite{Kiguchi08}.

\section{Results and Discussion}
%\label{sec:results}

\begin{figure*}
        \centering
        \subfloat[Full many-body MDE theory]{\label{fig:many-body_benzene}
        \includegraphics[width=.5\linewidth]{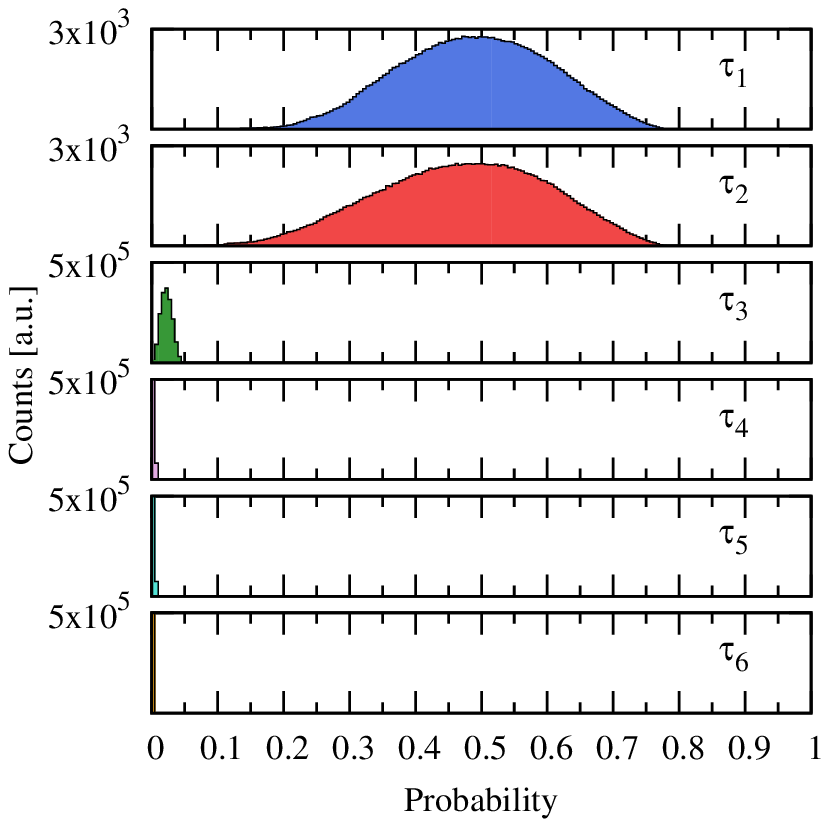}}
        \subfloat[Isolated ({\sc HOMO}) resonance approximation]{\label{fig:iso_res_benzene}
        \includegraphics[width=.5\linewidth]{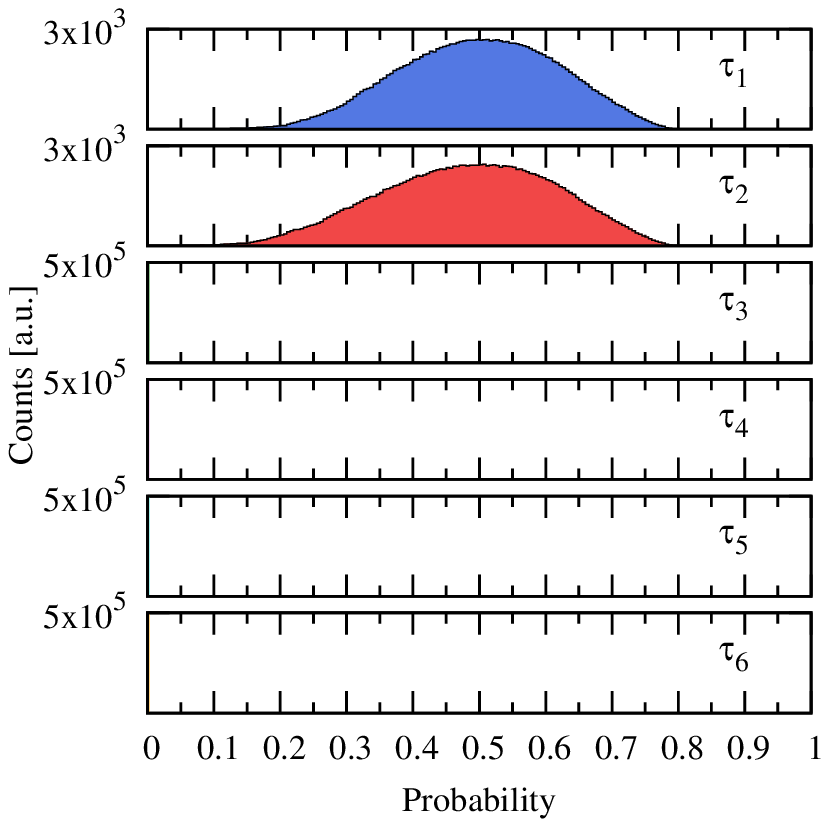}}
        \caption{The calculated eigenvalue distributions for an ensemble of $1.74\times 10^5$ (2000 bonding configuration $\times$ 87 Pt surfaces) Pt--benzene--Pt junctions using many-body theory with (a) the full spectrum and (b) the isolated-resonance approximation for the (doubly degenerate) \HO resonance.  Despite each lead forming $\sim 5$ bonds (cf. Fig.~\ref{fig:Gamma1_eigenchannel}), calculations in both cases exhibit only two dominant channels which arise from the degeneracy of the relevant ({\sc HOMO}) resonance.  The weak third channel seen in (a) is a consequence of the large lead-molecule coupling and is consistent with the measurements of Ref.\citenum{Kiguchi08}.}
        \label{fig:transport_results}
\end{figure*}

%In this section we present the transport using the aforementoined techniques.

%   In addition to sampling bonding configurations, we also consider the ensemble of junctions %produced in the experiment \cite{Kiguchi08}
%to sample all possible Pt surfaces.  The work function of Pt ranges from 5.93eV to 5.12eV for the (111) and (331) surfaces,
%respectively \cite{CRC}, and we assume that $\mu_{\rm Pt}$ is distributed uniformly over this interval.  

% Present fig 6a and 6b and say how well they correspond
The transmission eigenvalue distributions
for ensembles of $1.74\times 10^5$ Pt--benzene--Pt junctions calculated using the full many-body spectrum and in the isolated-resonance approximation are shown in %\ref{fig:many-body_benzene,fig:iso_res_benzene}
Figs.~\ref{fig:many-body_benzene} and \ref{fig:iso_res_benzene}, respectively.  
Despite the existence of five covalent bonds between the molecule and each lead (cf.\ Fig.~\ref{fig:Gamma1_eigenchannel}), 
there are only two dominant transmission channels, which arise from the two-fold degenerate \HO resonance closest to the Pt Fermi level \cite{Bergfield11a}.  
As proof of this point, we calculated the transmission eigenvalue distribution, over the same ensemble, using only the \HO resonance in the 
isolated-resonance approximation (Eq.\ref{eq:isoresapprox}).  
The resulting transmission eigenvalue distributions, 
shown in Fig.~\ref{fig:iso_res_benzene}, 
are nearly identical to the full %many-body 
distribution shown in 
Fig.~\ref{fig:many-body_benzene}, with the exception of the small 
but experimentally resolvable \cite{Kiguchi08} third transmission channel.

The lack of a third channel in the isolated-resonance approximation
is a direct consequence of the two-fold degeneracy
of the \HO resonance, which can therefore contribute at most two transmission channels.  The %origin of the 
third channel thus arises from 
further off-resonant tunneling.  
%when using the full molecular spectrum
In fact, we would argue that the very observation of a third channel in some Pt--benzene-Pt junctions \cite{Kiguchi08}
is a consequence of the very large lead-molecule coupling ($\sim 2eV$ per atomic orbital) in 
this system.  Having simulated junctions with electrodes whose DOS at the Fermi 
level is smaller than that of Pt, we expect junctions with Cu or Au electrodes, for example, to 
exhibit only two measurable transmission channels.

\begin{figure}
	\centering
		\includegraphics{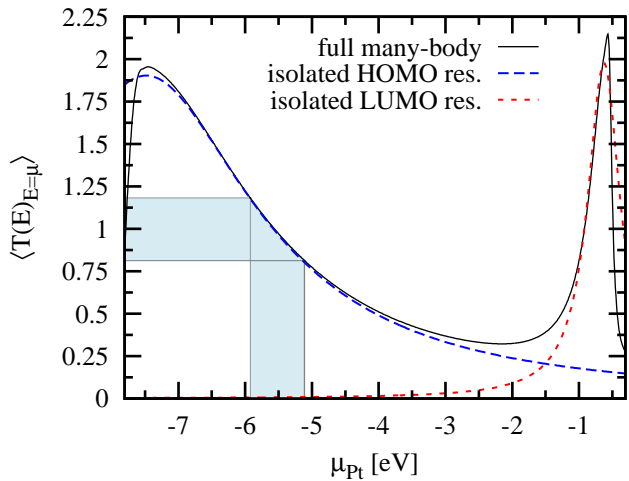}
	\caption{The calculated average total transmission averaged over 2000 bonding 
	configurations through a Pt--benzene--Pt junction shown as a function of 
	the leads' chemical potential $\mu_{\rm Pt}$.  The isolated resonance approximation using the \HO or \LU resonance accurately describe the full many-body transport in the vicinity of the \HO or \LU resonance, respectively.  These data are in good agreement with the measurements of Ref.~\citenum{Kiguchi08}.  %This spectrum lacks particle-hole symmetry because of the constraints placed on the generation of the ensemble.  (TODO: put this in text instead/also?) In a previous work the spectrum was symmetric \cite{Bergfield11a}. 
The work-function range for the crystal planes of Pt are shaded in blue, where $-5.84eV \leq \mu_{\rm Pt} \leq -5.12eV$ \cite{CRC}.}
	\label{fig:average_transmission}
\end{figure}

% Isolated resonance
%(TODO: work)

In order to investigate the efficacy of the isolated-resonance approximation further,
we calculated the average total transmission through a Pt--benzene--Pt junction.  
The transmission spectra calculated using the full molecular spectrum, the isolated \HO resonance %approximation
and the isolated \LU resonance %approximation 
are each shown as a function of the leads' chemical potential $\mu_{\rm Pt}$ in Fig.~\ref{fig:average_transmission}.
The spectra are averaged over 2000 bonding configurations and  
the blue shaded area indicates the range of possible chemical potentials for the Pt electrodes.  
The close correspondence between the full transmission spectrum and the isolated \HO resonance %spectra
over this range is consistent with the accuracy of the approximate method shown in Fig.~\ref{fig:transport_results}.  
Similarly, in the vicinity of the \LU resonance,
the isolated \LU resonance approximation accurately characterizes the average transmission.
The \HOLU asymmetry in the average transmission function arises because the
{\sc HOMO} resonance couples more strongly on average to the Pt tip atoms than does the {\sc LUMO} resonance.

%(TODO: keep?) 
%We mention that the asymmetry of the full many-body transmission spectrum, with respect to the \HOLU mid-gap ($\mu_{\rm Pt}=-4.055eV$), is a consequence of the physically motivated 
%restrictions built into the ensemble generation method.  The random walk method developed in Ref.~\citenum{Bergfield11a} had no such constraints and consequently predicted particle-hole symmetric transmission spectra.

It is tempting to assume, based on the accuracy of the isolated-resonance approximation
in our many-body transport theory,
that an analogous ``single molecular orbital'' approximation 
would also be sufficient in a transport calculation based e.g.\ on density-functional theory (DFT).
However, this is not the case.
Although the isolated-resonance approximation can also be derived
within DFT, in practice,
%the absolute squares of which can be interpreted as their contributions %of each molecular resonance 
%to each transmission channel.
it is necessary to use an ``extended molecule'' %must be used %\cite{Heurich02} 
to account for charge transfer between molecule and electrodes.  
This is because current implementations of DFT fail
to account for the {\em particle aspect} of the electron  \cite{Toher05,Burke06,Datta06,Geskin09}.  
%, {\em i.e.}, the %strong
%tendency for the electric charge on the molecule within the junction to be quantized in integer multiples of the electron charge $e$.
Analyzing transport in terms of extended molecular orbitals has proven problematic.
%in terms of the resonances of the molecule itself. % \cite{Heurich02}.
For example, the resonances of the extended molecule in Ref.\ \citenum{Heurich02} apparently accounted for less than 9\% of the current through the junction.

Using an ``extended molecule'' also makes it difficult, if not impossible,
to interpret transport contributions in terms of the resonances of the molecule itself \cite{Heurich02}.  
Since charging effects in SMJs are well-described in our many-body theory \cite{Bergfield09, Bergfield11b}, there is no need to utilize an 
``extended molecule,''
so the resonances in our isolated-resonance approximation are true molecular resonances.

\begin{figure}
	\centering
		\includegraphics{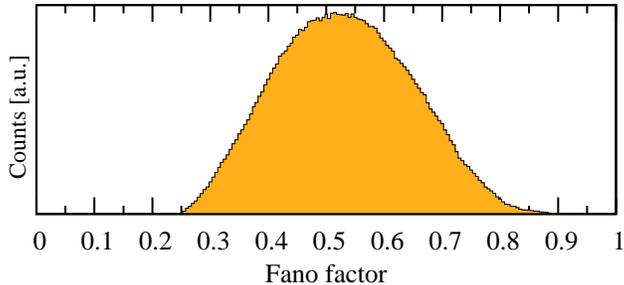}
	\caption{The calculated Fano factor $F$ distribution for the full ensemble of $1.74\times10^5$ Pt--benzene--Pt junctions.   $F$ describes the nature of the transport, where $F=0$ and $F=1$ characterize wave-like (ballistic) and particle-like transport, respectively.  The peak value of this distribution $F\sim 0.51$ indicates that we are in an intermediate regime.}
	\label{fig:fano_factor}
\end{figure}

The full counting statistics of a distribution are characterized by its cumulants.  Using a single-particle theory to describe a single-channel junction, it can be shown \cite{Levitov96,Levitov93} that the first cumulant is related to the junction transmission function while the second cumulant is related to the shot noise suppression.  Often this suppression is phrased in terms of the Fano factor \cite{Schottky18}
\begin{equation}
	F= \sum_n \frac{\tau_n (1-\tau_n)}{\sum_n \tau_n}.
\end{equation}
 In Fig.~\ref{fig:fano_factor} we show the distribution of $F$ for our ensemble of junctions, where the $\tau_n$ have been calculated using many-body theory.  Because of the fermionic character of the charge carriers $0 \leq F \leq 1$, with $F=0$ corresponding to completely wave-like transport and a value of $F=1$ corresponding to completely particle-like transport.  
From the figure, we see that $F$ is peaked $\sim 0.51$ implying that the {\em both particle and wave} aspects of the carriers are important, 
a fact which is consistent with the commensurate charging energy and bonding strength (cf.\  Fig.~\ref{fig:benzene_Unm_Trace_Gamma}).  

%(TODO: good enough?)
In such an intermediate regime both the `complementary' aspects of the charge carriers are equally important, requiring a many-body description and resulting in many subtle and interesting effects.  For example, the transport in this regime displays a variety of features stemming from the interplay between Coulomb blockade and coherent interference effects, which occur simultaneously \cite{Bergfield09,Bergfield10b}. 
 Although the Fano factor reflects the nature of the transport, it is not directly related to the shot-noise power in a many-body theory.
The richness of the transport %phenomena exhibited 
in this regime, however, suggests that a full many-body calculation of a higher-order moment, such as the shot-noise, may exhibit equally interesting phenomena. % that warrant further study.

\section{Conclusion}

% 
%In conclusion, 
We have developed a state-of-the-art technique to model the lead-molecule coupling in highly-conductive molecular junctions.
%which we then used to investigate transport in ensembles of Pt--benzene-Pt junctions.  
The 
bonding between the lead and molecule was described using an `ab initio' model in which the tunneling matrix elements between
all relevant lead tip wavefunctions and the molecule were included, producing multi-channel
 junctions naturally from a physically motivated ensemble over contact geometries.  Coulomb
interactions between the molecule and the metallic leads were included using an image multipole
method within $\pi$-EFT. % \cite{Barr11}.
%influence of the leads on the molecular $\pi$-electron system was described by an $\pi$-EFT, giving an
%effective Hamiltonian for the molecule without the necessity of resorting to 
%semi-empirical methods which accounts for intramolecular screening and van der Waals effects.
In concert, these techniques allowed us to accurately model SMJs within our many-body theory.

%
%interace 
%Using these models we were able to account for intramolecular screening and van der Waals effects 

% What we found
The transport
%Using the our model for the lead-molecule coupling, we calculated the transport 
for
an ensemble of Pt--benzene--Pt junctions, calculated using our many-body theory, confirmed our
statement \cite{Bergfield11a} that the number of dominant transmission
channels in a SMJ is equal to the degeneracy of the
molecular orbital closest to the metal Fermi level.  
We find that the transport through a Pt--benzene--Pt junction
can be accurately described using only the relevant ({\sc HOMO}) molecular resonance.
The exceptional accuracy of such an isolated-resonance approximation, however, may be limited to
small molecules with large charging energies.  
In larger molecules, where the charging energy is smaller, 
further off-resonant transmission channels are expected to become more important.

In metallic point contacts the number of channels
is completely determined by the valence of the metal.  
Despite the larger number of states available for
tunneling transport in SMJs, we predict that the number
of transmission channels is typically more limited
than in single-atom contacts because molecules are
less symmetrical than atoms.
Channel-resolved transport measurements of SMJs therefore offer a unique probe into the
symmetry of the molecular species involved.

\bibliography{refs}

\begin{thebibliography}{34}%
\makeatletter
\providecommand \@ifxundefined [1]{%
 \@ifx{#1\undefined}
}%
\providecommand \@ifnum [1]{%
 \ifnum #1\expandafter \@firstoftwo
 \else \expandafter \@secondoftwo
 \fi
}%
\providecommand \@ifx [1]{%
 \ifx #1\expandafter \@firstoftwo
 \else \expandafter \@secondoftwo
 \fi
}%
\providecommand \natexlab [1]{#1}%
\providecommand \enquote  [1]{``#1''}%
\providecommand \bibnamefont  [1]{#1}%
\providecommand \bibfnamefont [1]{#1}%
\providecommand \citenamefont [1]{#1}%
\providecommand \href@noop [0]{\@secondoftwo}%
\providecommand \href [0]{\begingroup \@sanitize@url \@href}%
\providecommand \@href[1]{\@@startlink{#1}\@@href}%
\providecommand \@@href[1]{\endgroup#1\@@endlink}%
\providecommand \@sanitize@url [0]{\catcode `\\12\catcode `\$12\catcode
  `\&12\catcode `\#12\catcode `\^12\catcode `\_12\catcode `\%12\relax}%
\providecommand \@@startlink[1]{}%
\providecommand \@@endlink[0]{}%
\providecommand \url  [0]{\begingroup\@sanitize@url \@url }%
\providecommand \@url [1]{\endgroup\@href {#1}{\urlprefix }}%
\providecommand \urlprefix  [0]{URL }%
\providecommand \Eprint [0]{\href }%
\providecommand \doibase [0]{http://dx.doi.org/}%
\providecommand \selectlanguage [0]{\@gobble}%
\providecommand \bibinfo  [0]{\@secondoftwo}%
\providecommand \bibfield  [0]{\@secondoftwo}%
\providecommand \translation [1]{[#1]}%
\providecommand \BibitemOpen [0]{}%
\providecommand \bibitemStop [0]{}%
\providecommand \bibitemNoStop [0]{.\EOS\space}%
\providecommand \EOS [0]{\spacefactor3000\relax}%
\providecommand \BibitemShut  [1]{\csname bibitem#1\endcsname}%
\let\auto@bib@innerbib\@empty
%</preamble>
\bibitem [{\citenamefont {Scheer}\ \emph {et~al.}(1998)\citenamefont {Scheer},
  \citenamefont {{Agra\"{\i}t}}, \citenamefont {Cuevas}, \citenamefont {{Levy
  Yeyati}}, \citenamefont {Ludoph}, \citenamefont {Mart{\'\i}n-Rodero},
  \citenamefont {{Rubio Bollinger}}, \citenamefont {van Ruitenbeek},\ and\
  \citenamefont {Urbina}}]{Scheer98}%
  \BibitemOpen
  \bibfield  {author} {\bibinfo {author} {\bibfnamefont {E.}~\bibnamefont
  {Scheer}}, \bibinfo {author} {\bibfnamefont {N.}~\bibnamefont
  {{Agra\"{\i}t}}}, \bibinfo {author} {\bibfnamefont {J.~C.}\ \bibnamefont
  {Cuevas}}, \bibinfo {author} {\bibfnamefont {A.}~\bibnamefont {{Levy
  Yeyati}}}, \bibinfo {author} {\bibfnamefont {B.}~\bibnamefont {Ludoph}},
  \bibinfo {author} {\bibfnamefont {A.}~\bibnamefont {Mart{\'\i}n-Rodero}},
  \bibinfo {author} {\bibfnamefont {G.}~\bibnamefont {{Rubio Bollinger}}},
  \bibinfo {author} {\bibfnamefont {J.~M.}\ \bibnamefont {van Ruitenbeek}}, \
  and\ \bibinfo {author} {\bibfnamefont {C.}~\bibnamefont {Urbina}},\
  }\href@noop {} {\bibfield  {journal} {\bibinfo  {journal} {Nature}\ }\textbf
  {\bibinfo {volume} {394}},\ \bibinfo {pages} {154} (\bibinfo {year}
  {1998})}\BibitemShut {NoStop}%
\bibitem [{\citenamefont {Bergfield}\ \emph
  {et~al.}(2011{\natexlab{a}})\citenamefont {Bergfield}, \citenamefont {Barr},\
  and\ \citenamefont {Stafford}}]{Bergfield11a}%
  \BibitemOpen
  \bibfield  {author} {\bibinfo {author} {\bibfnamefont {J.~P.}\ \bibnamefont
  {Bergfield}}, \bibinfo {author} {\bibfnamefont {J.~D.}\ \bibnamefont {Barr}},
  \ and\ \bibinfo {author} {\bibfnamefont {C.~A.}\ \bibnamefont {Stafford}},\
  }\href {http://pubs.acs.org/doi/abs/10.1021/nn1030753} {\bibfield  {journal}
  {\bibinfo  {journal} {ACS Nano}\ }\textbf {\bibinfo {volume} {5}},\ \bibinfo
  {pages} {2707} (\bibinfo {year} {2011}{\natexlab{a}})}\BibitemShut {NoStop}%
\bibitem [{foo()}]{footnote_HOMOLUMO}%
  \BibitemOpen
  \href@noop {} {\ }\bibinfo {note} {``Molecular orbitals'' are a
  single-particle concept and not defined in a many-body theory. Instead we
  define the ``\HO resonance'' as the addition spectrum peak corresponding to
  the $N-1 \rightarrow N$ electronic transition of the molecule, where $N$ is
  the charge of the neutral molecule. Similarly, we define the ``\LU
  resonance'' as the addition spectrum peak corresponding to the $N \rightarrow
  N+1$ molecular charge transition.}\BibitemShut {Stop}%
\bibitem [{\citenamefont {Kiguchi}\ \emph {et~al.}(2008)\citenamefont
  {Kiguchi}, \citenamefont {Tal}, \citenamefont {Wohlthat}, \citenamefont
  {Pauly}, \citenamefont {Krieger}, \citenamefont {Djukic}, \citenamefont
  {Cuevas},\ and\ \citenamefont {van Ruitenbeek}}]{Kiguchi08}%
  \BibitemOpen
  \bibfield  {author} {\bibinfo {author} {\bibfnamefont {M.}~\bibnamefont
  {Kiguchi}}, \bibinfo {author} {\bibfnamefont {O.}~\bibnamefont {Tal}},
  \bibinfo {author} {\bibfnamefont {S.}~\bibnamefont {Wohlthat}}, \bibinfo
  {author} {\bibfnamefont {F.}~\bibnamefont {Pauly}}, \bibinfo {author}
  {\bibfnamefont {M.}~\bibnamefont {Krieger}}, \bibinfo {author} {\bibfnamefont
  {D.}~\bibnamefont {Djukic}}, \bibinfo {author} {\bibfnamefont {J.~C.}\
  \bibnamefont {Cuevas}}, \ and\ \bibinfo {author} {\bibfnamefont {J.~M.}\
  \bibnamefont {van Ruitenbeek}},\ }\href@noop {} {\bibfield  {journal}
  {\bibinfo  {journal} {Phys. Rev. Lett.}\ }\textbf {\bibinfo {volume} {101}}
  (\bibinfo {year} {2008})}\BibitemShut {NoStop}%
\bibitem [{\citenamefont {Datta}(1995)}]{Datta95}%
  \BibitemOpen
  \bibfield  {author} {\bibinfo {author} {\bibfnamefont {S.}~\bibnamefont
  {Datta}},\ }in\ \href@noop {} {\emph {\bibinfo {booktitle} {Electronic
  Transport in Mesoscopic Systems}}}\ (\bibinfo  {publisher} {Cambridge
  University Press},\ \bibinfo {address} {Cambridge, UK},\ \bibinfo {year}
  {1995})\ pp.\ \bibinfo {pages} {117--174}\BibitemShut {NoStop}%
\bibitem [{\citenamefont {Bergfield}\ and\ \citenamefont
  {Stafford}(2009)}]{Bergfield09}%
  \BibitemOpen
  \bibfield  {author} {\bibinfo {author} {\bibfnamefont {J.~P.}\ \bibnamefont
  {Bergfield}}\ and\ \bibinfo {author} {\bibfnamefont {C.~A.}\ \bibnamefont
  {Stafford}},\ }\href@noop {} {\bibfield  {journal} {\bibinfo  {journal}
  {Phys. Rev. B}\ }\textbf {\bibinfo {volume} {79}},\ \bibinfo {pages} {245125}
  (\bibinfo {year} {2009})}\BibitemShut {NoStop}%
\bibitem [{\citenamefont {Toher}\ \emph {et~al.}(2005)\citenamefont {Toher},
  \citenamefont {Filippetti}, \citenamefont {Sanvito},\ and\ \citenamefont
  {Burke}}]{Toher05}%
  \BibitemOpen
  \bibfield  {author} {\bibinfo {author} {\bibfnamefont {C.}~\bibnamefont
  {Toher}}, \bibinfo {author} {\bibfnamefont {A.}~\bibnamefont {Filippetti}},
  \bibinfo {author} {\bibfnamefont {S.}~\bibnamefont {Sanvito}}, \ and\
  \bibinfo {author} {\bibfnamefont {K.}~\bibnamefont {Burke}},\ }\href@noop {}
  {\bibfield  {journal} {\bibinfo  {journal} {Phys. Rev. Lett.}\ }\textbf
  {\bibinfo {volume} {95}},\ \bibinfo {pages} {146402} (\bibinfo {year}
  {2005})}\BibitemShut {NoStop}%
\bibitem [{\citenamefont {Koentopp}\ \emph {et~al.}(2006)\citenamefont
  {Koentopp}, \citenamefont {Burke},\ and\ \citenamefont {Evers}}]{Burke06}%
  \BibitemOpen
  \bibfield  {author} {\bibinfo {author} {\bibfnamefont {M.}~\bibnamefont
  {Koentopp}}, \bibinfo {author} {\bibfnamefont {K.}~\bibnamefont {Burke}}, \
  and\ \bibinfo {author} {\bibfnamefont {F.}~\bibnamefont {Evers}},\
  }\href@noop {} {\bibfield  {journal} {\bibinfo  {journal} {Phys. Rev. B}\
  }\textbf {\bibinfo {volume} {73}},\ \bibinfo {pages} {121403} (\bibinfo
  {year} {2006})}\BibitemShut {NoStop}%
\bibitem [{\citenamefont {Muralidharan}\ \emph {et~al.}(2006)\citenamefont
  {Muralidharan}, \citenamefont {Ghosh},\ and\ \citenamefont
  {Datta}}]{Datta06}%
  \BibitemOpen
  \bibfield  {author} {\bibinfo {author} {\bibfnamefont {B.}~\bibnamefont
  {Muralidharan}}, \bibinfo {author} {\bibfnamefont {A.~W.}\ \bibnamefont
  {Ghosh}}, \ and\ \bibinfo {author} {\bibfnamefont {S.}~\bibnamefont
  {Datta}},\ }\href@noop {} {\bibfield  {journal} {\bibinfo  {journal} {Phys.
  Rev. B}\ }\textbf {\bibinfo {volume} {73}},\ \bibinfo {pages} {155410}
  (\bibinfo {year} {2006})}\BibitemShut {NoStop}%
\bibitem [{\citenamefont {Geskin}\ \emph {et~al.}(2009)\citenamefont {Geskin},
  \citenamefont {Stadler},\ and\ \citenamefont {Cornil}}]{Geskin09}%
  \BibitemOpen
  \bibfield  {author} {\bibinfo {author} {\bibfnamefont {V.}~\bibnamefont
  {Geskin}}, \bibinfo {author} {\bibfnamefont {R.}~\bibnamefont {Stadler}}, \
  and\ \bibinfo {author} {\bibfnamefont {J.}~\bibnamefont {Cornil}},\
  }\href@noop {} {\bibfield  {journal} {\bibinfo  {journal} {Phys. Rev. B}\
  }\textbf {\bibinfo {volume} {80}},\ \bibinfo {pages} {085411} (\bibinfo
  {year} {2009})}\BibitemShut {NoStop}%
\bibitem [{\citenamefont {Bergfield}\ \emph {et~al.}(2010)\citenamefont
  {Bergfield}, \citenamefont {Jacquod},\ and\ \citenamefont
  {Stafford}}]{Bergfield10b}%
  \BibitemOpen
  \bibfield  {author} {\bibinfo {author} {\bibfnamefont {J.~P.}\ \bibnamefont
  {Bergfield}}, \bibinfo {author} {\bibfnamefont {P.}~\bibnamefont {Jacquod}},
  \ and\ \bibinfo {author} {\bibfnamefont {C.~A.}\ \bibnamefont {Stafford}},\
  }\href@noop {} {\bibfield  {journal} {\bibinfo  {journal} {Phys. Rev. B}\
  }\textbf {\bibinfo {volume} {82}},\ \bibinfo {pages} {205405} (\bibinfo
  {year} {2010})}\BibitemShut {NoStop}%
\bibitem [{\citenamefont {Bergfield}\ \emph
  {et~al.}(2011{\natexlab{b}})\citenamefont {Bergfield}, \citenamefont
  {Solomon}, \citenamefont {Stafford},\ and\ \citenamefont
  {Ratner}}]{Bergfield11b}%
  \BibitemOpen
  \bibfield  {author} {\bibinfo {author} {\bibfnamefont {J.~P.}\ \bibnamefont
  {Bergfield}}, \bibinfo {author} {\bibfnamefont {G.~C.}\ \bibnamefont
  {Solomon}}, \bibinfo {author} {\bibfnamefont {C.~A.}\ \bibnamefont
  {Stafford}}, \ and\ \bibinfo {author} {\bibfnamefont {M.~A.}\ \bibnamefont
  {Ratner}},\ }\href {http://pubs.acs.org/doi/abs/10.1021/nl201042m} {\bibfield
   {journal} {\bibinfo  {journal} {Nano Letters}\ }\textbf {\bibinfo {volume}
  {11}},\ \bibinfo {pages} {2759} (\bibinfo {year}
  {2011}{\natexlab{b}})}\BibitemShut {NoStop}%
\bibitem [{\citenamefont {Castleton}\ and\ \citenamefont
  {Barford}(2002)}]{Castleton02}%
  \BibitemOpen
  \bibfield  {author} {\bibinfo {author} {\bibfnamefont {C.~W.~M.}\
  \bibnamefont {Castleton}}\ and\ \bibinfo {author} {\bibfnamefont
  {W.}~\bibnamefont {Barford}},\ }\href
  {http://link.aip.org/link/?JCP/117/3570/1} {\bibfield  {journal} {\bibinfo
  {journal} {J. Chem. Phys.}\ }\textbf {\bibinfo {volume} {117}},\ \bibinfo
  {pages} {3570} (\bibinfo {year} {2002})}\BibitemShut {NoStop}%
\bibitem [{\citenamefont {Djukic}\ and\ \citenamefont {van
  Ruitenbeek}(2006)}]{Djukic06}%
  \BibitemOpen
  \bibfield  {author} {\bibinfo {author} {\bibfnamefont {D.}~\bibnamefont
  {Djukic}}\ and\ \bibinfo {author} {\bibfnamefont {J.~M.}\ \bibnamefont {van
  Ruitenbeek}},\ }\href@noop {} {\bibfield  {journal} {\bibinfo  {journal}
  {Nano Lett.}\ }\textbf {\bibinfo {volume} {6}},\ \bibinfo {pages} {789}
  (\bibinfo {year} {2006})}\BibitemShut {NoStop}%
\bibitem [{\citenamefont {Solomon}\ \emph {et~al.}(2006)\citenamefont
  {Solomon}, \citenamefont {Gagliardi}, \citenamefont {Pecchia}, \citenamefont
  {Frauenheim}, \citenamefont {Di~Carlo}, \citenamefont {Reimers},\ and\
  \citenamefont {Hush}}]{Solomon06b}%
  \BibitemOpen
  \bibfield  {author} {\bibinfo {author} {\bibfnamefont {G.~C.}\ \bibnamefont
  {Solomon}}, \bibinfo {author} {\bibfnamefont {A.}~\bibnamefont {Gagliardi}},
  \bibinfo {author} {\bibfnamefont {A.}~\bibnamefont {Pecchia}}, \bibinfo
  {author} {\bibfnamefont {T.}~\bibnamefont {Frauenheim}}, \bibinfo {author}
  {\bibfnamefont {A.}~\bibnamefont {Di~Carlo}}, \bibinfo {author}
  {\bibfnamefont {J.~R.}\ \bibnamefont {Reimers}}, \ and\ \bibinfo {author}
  {\bibfnamefont {N.~S.}\ \bibnamefont {Hush}},\ }\href@noop {} {\bibfield
  {journal} {\bibinfo  {journal} {Nano Lett.}\ }\textbf {\bibinfo {volume}
  {6}},\ \bibinfo {pages} {2431} (\bibinfo {year} {2006})}\BibitemShut
  {NoStop}%
\bibitem [{\citenamefont {Barr}\ \emph {et~al.}(2011)\citenamefont {Barr},
  \citenamefont {Bergfield},\ and\ \citenamefont {Stafford}}]{Barr11}%
  \BibitemOpen
  \bibfield  {author} {\bibinfo {author} {\bibfnamefont {J.~D.}\ \bibnamefont
  {Barr}}, \bibinfo {author} {\bibfnamefont {J.~P.}\ \bibnamefont {Bergfield}},
  \ and\ \bibinfo {author} {\bibfnamefont {C.~A.}\ \bibnamefont {Stafford}},\
  }\href@noop {} {\  (\bibinfo {year} {2011})},\ \bibinfo {note}
  {unpublished}\BibitemShut {NoStop}%
\bibitem [{\citenamefont {Kovac}\ \emph {et~al.}(1980)\citenamefont {Kovac},
  \citenamefont {Mohraz}, \citenamefont {Heilbronner}, \citenamefont
  {Boekelheide},\ and\ \citenamefont {Hopf}}]{Kovac80}%
  \BibitemOpen
  \bibfield  {author} {\bibinfo {author} {\bibfnamefont {B.}~\bibnamefont
  {Kovac}}, \bibinfo {author} {\bibfnamefont {M.}~\bibnamefont {Mohraz}},
  \bibinfo {author} {\bibfnamefont {E.}~\bibnamefont {Heilbronner}}, \bibinfo
  {author} {\bibfnamefont {V.}~\bibnamefont {Boekelheide}}, \ and\ \bibinfo
  {author} {\bibfnamefont {H.}~\bibnamefont {Hopf}},\ }\href {\doibase
  10.1021/ja00533a005} {\bibfield  {journal} {\bibinfo  {journal} {J. Am. Chem.
  Soc.}\ }\textbf {\bibinfo {volume} {102}},\ \bibinfo {pages} {4314} (\bibinfo
  {year} {1980})}\BibitemShut {NoStop}%
\bibitem [{\citenamefont {Sell}\ and\ \citenamefont
  {Kuppermann}(1978)}]{Sell78}%
  \BibitemOpen
  \bibfield  {author} {\bibinfo {author} {\bibfnamefont {J.~A.}\ \bibnamefont
  {Sell}}\ and\ \bibinfo {author} {\bibfnamefont {A.}~\bibnamefont
  {Kuppermann}},\ }\href
  {http://www.sciencedirect.com/science/article/B6TFM-44XDY26-BH/2/10ad19dc6dd545479060cc3e7e63dec2}
  {\bibfield  {journal} {\bibinfo  {journal} {Chemical Physics}\ }\textbf
  {\bibinfo {volume} {33}},\ \bibinfo {pages} {367 } (\bibinfo {year}
  {1978})}\BibitemShut {NoStop}%
\bibitem [{\citenamefont {Kobayoshi}(1978)}]{Kobayoshi78}%
  \BibitemOpen
  \bibfield  {author} {\bibinfo {author} {\bibfnamefont {T.}~\bibnamefont
  {Kobayoshi}},\ }\href
  {http://www.sciencedirect.com/science/article/B6TVM-46SPKFJ-JT/2/21469c391855faa3e978b3ac29f5355c}
  {\bibfield  {journal} {\bibinfo  {journal} {Phys. Rev. A}\ }\textbf {\bibinfo
  {volume} {69}},\ \bibinfo {pages} {105 } (\bibinfo {year}
  {1978})}\BibitemShut {NoStop}%
\bibitem [{\citenamefont {Schmidt}(1977)}]{Schmidt77}%
  \BibitemOpen
  \bibfield  {author} {\bibinfo {author} {\bibfnamefont {W.}~\bibnamefont
  {Schmidt}},\ }\href {http://link.aip.org/link/?JCP/66/828/1} {\bibfield
  {journal} {\bibinfo  {journal} {J. Chem. Phys.}\ }\textbf {\bibinfo {volume}
  {66}},\ \bibinfo {pages} {828} (\bibinfo {year} {1977})}\BibitemShut
  {NoStop}%
\bibitem [{\citenamefont {Baltzer}\ \emph {et~al.}(1997)\citenamefont
  {Baltzer}, \citenamefont {Karlsson}, \citenamefont {Wannberg}, \citenamefont
  {\"Ohrwall}, \citenamefont {Holland}, \citenamefont {MacDonald},
  \citenamefont {Hayes},\ and\ \citenamefont {von Niessen}}]{Baltzer97}%
  \BibitemOpen
  \bibfield  {author} {\bibinfo {author} {\bibfnamefont {P.}~\bibnamefont
  {Baltzer}}, \bibinfo {author} {\bibfnamefont {L.}~\bibnamefont {Karlsson}},
  \bibinfo {author} {\bibfnamefont {B.}~\bibnamefont {Wannberg}}, \bibinfo
  {author} {\bibfnamefont {G.}~\bibnamefont {\"Ohrwall}}, \bibinfo {author}
  {\bibfnamefont {D.~M.~P.}\ \bibnamefont {Holland}}, \bibinfo {author}
  {\bibfnamefont {M.~A.}\ \bibnamefont {MacDonald}}, \bibinfo {author}
  {\bibfnamefont {M.~A.}\ \bibnamefont {Hayes}}, \ and\ \bibinfo {author}
  {\bibfnamefont {W.}~\bibnamefont {von Niessen}},\ }\href
  {http://www.sciencedirect.com/science/article/B6TFM-3SFVB7R-15/2/4f01f84ddc2fe09385e8957dbde6f029}
  {\bibfield  {journal} {\bibinfo  {journal} {Chemical Physics}\ }\textbf
  {\bibinfo {volume} {224}},\ \bibinfo {pages} {95 } (\bibinfo {year}
  {1997})}\BibitemShut {NoStop}%
\bibitem [{\citenamefont {Howell}\ \emph {et~al.}(1984)\citenamefont {Howell},
  \citenamefont {Goncalves}, \citenamefont {Amatore}, \citenamefont {Klasinc},
  \citenamefont {Wightman},\ and\ \citenamefont {Kochi}}]{Howell84}%
  \BibitemOpen
  \bibfield  {author} {\bibinfo {author} {\bibfnamefont {J.~O.}\ \bibnamefont
  {Howell}}, \bibinfo {author} {\bibfnamefont {J.~M.}\ \bibnamefont
  {Goncalves}}, \bibinfo {author} {\bibfnamefont {C.}~\bibnamefont {Amatore}},
  \bibinfo {author} {\bibfnamefont {L.}~\bibnamefont {Klasinc}}, \bibinfo
  {author} {\bibfnamefont {R.~M.}\ \bibnamefont {Wightman}}, \ and\ \bibinfo
  {author} {\bibfnamefont {J.~K.}\ \bibnamefont {Kochi}},\ }\href
  {http://pubs.acs.org/doi/abs/10.1021/ja00326a014} {\bibfield  {journal}
  {\bibinfo  {journal} {J. Am. Chem. Soc.}\ }\textbf {\bibinfo {volume}
  {106}},\ \bibinfo {pages} {3968} (\bibinfo {year} {1984})}\BibitemShut
  {NoStop}%
\bibitem [{\citenamefont {Burrow}\ \emph {et~al.}(1987)\citenamefont {Burrow},
  \citenamefont {Michejda},\ and\ \citenamefont {Jordan}}]{Burrow87}%
  \BibitemOpen
  \bibfield  {author} {\bibinfo {author} {\bibfnamefont {P.~D.}\ \bibnamefont
  {Burrow}}, \bibinfo {author} {\bibfnamefont {J.~A.}\ \bibnamefont
  {Michejda}}, \ and\ \bibinfo {author} {\bibfnamefont {K.~D.}\ \bibnamefont
  {Jordan}},\ }\href {http://link.aip.org/link/?JCP/86/9/1} {\bibfield
  {journal} {\bibinfo  {journal} {J. Chem. Phys.}\ }\textbf {\bibinfo {volume}
  {86}},\ \bibinfo {pages} {9} (\bibinfo {year} {1987})}\BibitemShut {NoStop}%
\bibitem [{\citenamefont {Lide~{\it et al.}}(2005)}]{CRC}%
  \BibitemOpen
  \bibinfo {editor} {\bibfnamefont {D.~R.}\ \bibnamefont {Lide~{\it et al.}}},\
  ed.,\ \href@noop {} {\emph {\bibinfo {title} {CRC Handbook of Chemistry and
  Physics}}}\ (\bibinfo  {publisher} {CRC Press},\ \bibinfo {address} {Boca
  Raton, Fla.},\ \bibinfo {year} {2005})\BibitemShut {NoStop}%
\bibitem [{\citenamefont {Chen}(1993)}]{Chen93}%
  \BibitemOpen
  \bibfield  {author} {\bibinfo {author} {\bibfnamefont {C.~J.}\ \bibnamefont
  {Chen}},\ }\href@noop {} {\emph {\bibinfo {title} {Introduction to Scanning
  Tunneling Microscopy}}},\ \bibinfo {edition} {2nd}\ ed.\ (\bibinfo
  {publisher} {Oxford University Press, New York},\ \bibinfo {year}
  {1993})\BibitemShut {NoStop}%
\bibitem [{\citenamefont {Kleber}(1973)}]{Kleber73}%
  \BibitemOpen
  \bibfield  {author} {\bibinfo {author} {\bibfnamefont {R.}~\bibnamefont
  {Kleber}},\ }\href {http://dx.doi.org/10.1007/BF01398856} {\bibfield
  {journal} {\bibinfo  {journal} {Z. Phys. A: Hadrons Nucl.}\ }\textbf
  {\bibinfo {volume} {264}},\ \bibinfo {pages} {301} (\bibinfo {year}
  {1973})}\BibitemShut {NoStop}%
\bibitem [{\citenamefont {{Kittel}}(1976)}]{Kittel_KinderBook}%
  \BibitemOpen
  \bibfield  {author} {\bibinfo {author} {\bibfnamefont {C.}~\bibnamefont
  {{Kittel}}},\ }\href@noop {} {\emph {\bibinfo {title} {New York: Wiley, 1976,
  5th ed.}}}\ (\bibinfo  {publisher} {John Wiley and Sons, Inc.},\ \bibinfo
  {year} {1976})\BibitemShut {NoStop}%
\bibitem [{\citenamefont {Cruz}\ \emph {et~al.}(2007)\citenamefont {Cruz},
  \citenamefont {Carneiro}, \citenamefont {Aranda},\ and\ \citenamefont
  {B\~ahl}}]{Cruz07}%
  \BibitemOpen
  \bibfield  {author} {\bibinfo {author} {\bibfnamefont {M.~T. d.~M.}\
  \bibnamefont {Cruz}}, \bibinfo {author} {\bibfnamefont {J.~W. d.~M.}\
  \bibnamefont {Carneiro}}, \bibinfo {author} {\bibfnamefont {D.~A.~G.}\
  \bibnamefont {Aranda}}, \ and\ \bibinfo {author} {\bibfnamefont
  {M.}~\bibnamefont {B\~ahl}},\ }\href {http://dx.doi.org/10.1021/jp072572c}
  {\bibfield  {journal} {\bibinfo  {journal} {J. Phys. Chem. C}\ }\textbf
  {\bibinfo {volume} {111}},\ \bibinfo {pages} {11068} (\bibinfo {year}
  {2007})}\BibitemShut {NoStop}%
\bibitem [{\citenamefont {Morin}\ \emph {et~al.}(2003)\citenamefont {Morin},
  \citenamefont {Simon},\ and\ \citenamefont {Sautet}}]{Morin03}%
  \BibitemOpen
  \bibfield  {author} {\bibinfo {author} {\bibfnamefont {C.}~\bibnamefont
  {Morin}}, \bibinfo {author} {\bibfnamefont {D.}~\bibnamefont {Simon}}, \ and\
  \bibinfo {author} {\bibfnamefont {P.}~\bibnamefont {Sautet}},\ }\href
  {http://dx.doi.org/10.1021/jp026950j} {\bibfield  {journal} {\bibinfo
  {journal} {J. Phys. Chem. B}\ }\textbf {\bibinfo {volume} {107}},\ \bibinfo
  {pages} {2995} (\bibinfo {year} {2003})}\BibitemShut {NoStop}%
\bibitem [{\citenamefont {Saeys}\ \emph {et~al.}(2002)\citenamefont {Saeys},
  \citenamefont {Reyniers}, \citenamefont {Marin},\ and\ \citenamefont
  {Neurock}}]{Saeys02}%
  \BibitemOpen
  \bibfield  {author} {\bibinfo {author} {\bibfnamefont {M.}~\bibnamefont
  {Saeys}}, \bibinfo {author} {\bibfnamefont {M.-F.}\ \bibnamefont {Reyniers}},
  \bibinfo {author} {\bibfnamefont {G.~B.}\ \bibnamefont {Marin}}, \ and\
  \bibinfo {author} {\bibfnamefont {M.}~\bibnamefont {Neurock}},\ }\href
  {http://dx.doi.org/10.1021/jp0201231} {\bibfield  {journal} {\bibinfo
  {journal} {J. Phys. Chem. B}\ }\textbf {\bibinfo {volume} {106}},\ \bibinfo
  {pages} {7489} (\bibinfo {year} {2002})}\BibitemShut {NoStop}%
\bibitem [{\citenamefont {Heurich}\ \emph {et~al.}(2002)\citenamefont
  {Heurich}, \citenamefont {Cuevas}, \citenamefont {Wenzel},\ and\
  \citenamefont {Sch\"on}}]{Heurich02}%
  \BibitemOpen
  \bibfield  {author} {\bibinfo {author} {\bibfnamefont {J.}~\bibnamefont
  {Heurich}}, \bibinfo {author} {\bibfnamefont {J.~C.}\ \bibnamefont {Cuevas}},
  \bibinfo {author} {\bibfnamefont {W.}~\bibnamefont {Wenzel}}, \ and\ \bibinfo
  {author} {\bibfnamefont {G.}~\bibnamefont {Sch\"on}},\ }\href@noop {}
  {\bibfield  {journal} {\bibinfo  {journal} {Phys. Rev. Lett.}\ }\textbf
  {\bibinfo {volume} {88}},\ \bibinfo {pages} {256803} (\bibinfo {year}
  {2002})}\BibitemShut {NoStop}%
\bibitem [{\citenamefont {Levitov}\ \emph {et~al.}(1996)\citenamefont
  {Levitov}, \citenamefont {Lee},\ and\ \citenamefont {Lesovik}}]{Levitov96}%
  \BibitemOpen
  \bibfield  {author} {\bibinfo {author} {\bibfnamefont {L.~S.}\ \bibnamefont
  {Levitov}}, \bibinfo {author} {\bibfnamefont {H.}~\bibnamefont {Lee}}, \ and\
  \bibinfo {author} {\bibfnamefont {G.~B.}\ \bibnamefont {Lesovik}},\ }\href
  {http://dx.doi.org/doi/10.1063/1.531672} {\ \textbf {\bibinfo {volume}
  {37}},\ \bibinfo {pages} {4845} (\bibinfo {year} {1996})}\BibitemShut
  {NoStop}%
\bibitem [{\citenamefont {{Levitov}}\ and\ \citenamefont
  {{Lesovik}}(1993)}]{Levitov93}%
  \BibitemOpen
  \bibfield  {author} {\bibinfo {author} {\bibfnamefont {L.~S.}\ \bibnamefont
  {{Levitov}}}\ and\ \bibinfo {author} {\bibfnamefont {G.~B.}\ \bibnamefont
  {{Lesovik}}},\ }\href@noop {} {\bibfield  {journal} {\bibinfo  {journal}
  {JETP Letters}\ }\textbf {\bibinfo {volume} {58}},\ \bibinfo {pages} {230}
  (\bibinfo {year} {1993})}\BibitemShut {NoStop}%
\bibitem [{\citenamefont {Schottky}(1918)}]{Schottky18}%
  \BibitemOpen
  \bibfield  {author} {\bibinfo {author} {\bibfnamefont {W.}~\bibnamefont
  {Schottky}},\ }\href {http://dx.doi2.org/10.1002/andp.19183622304} {\bibfield
   {journal} {\bibinfo  {journal} {Annalen der Physik}\ }\textbf {\bibinfo
  {volume} {362}},\ \bibinfo {pages} {541} (\bibinfo {year}
  {1918})}\BibitemShut {NoStop}%
\end{thebibliography}%

%%%%%%%%%%%%%%%%%%%%%%%%%%%%%%%%%%%%%%%%%%%%%%%%%%%%%%%%%%%%%%%%%%%%%
%% That's it. Ending the document finishes the article. Happy TeXing!
%%%%%%%%%%%%%%%%%%%%%%%%%%%%%%%%%%%%%%%%%%%%%%%%%%%%%%%%%%%%%%%%%%%%%
\end{document}